\documentclass[lettersize,journal]{IEEEtran}
\usepackage{enumitem}
\usepackage{algorithm}
\usepackage{algpseudocode}
\usepackage{amsmath, amsfonts, amsthm, amssymb}
\usepackage{array}
\usepackage{stfloats}
\usepackage{textcomp}
\usepackage{stackengine}
\usepackage{url}
\usepackage{float} 
\usepackage{verbatim}
\usepackage{graphicx}
\usepackage{threeparttable}
\usepackage{subcaption}
\usepackage{xcolor}
\newtheorem{opportunity}{Opportunity}
\usepackage{tcolorbox}
\newtcolorbox{oppbox}{
    colback=gray!10!white, 
    colframe=black!70!white, 
    boxrule=0.5pt, 
    arc=2pt, 
    left=4pt, right=4pt, top=4pt, bottom=4pt, 
    fontupper=\small\itshape 
}

\usepackage[
    backend=biber,       
    style=ieee,       
    citestyle=numeric-comp, 
    maxcitenames=2,      
    maxbibnames=10,      
    minbibnames=10,   
    sorting=none         
]{biblatex}
\renewbibmacro*{url+urldate}{}
\AtEveryBibitem{%
  \clearfield{doi}
  \clearfield{isbn}
  \clearfield{location}
  \clearfield{month}
  \clearfield{note}
}

\usepackage{hyperref}
\usepackage{booktabs}
\usepackage{supertabular} 
\usepackage{multirow}
\usepackage{afterpage}  
\usepackage{cuted}

\begin{document}

\title{Beyond Traffic Matrix: DELTA – A DAG-Aware OCS Logical Topology Optimization for AIDCs}

\author{
    Niangen Ye,
    Jingya Liu,
    Guofu Zhu,
    Weiqiang Sun, \IEEEmembership{Senior Member,~IEEE},
    and Weisheng Hu, \textit{Member,~IEEE}
    \thanks{N. Ye, J. Liu, W. Sun, and W. Hu are with the State Key Laboratory of Photonics and Communications, Shanghai Jiao Tong University, Shanghai, China.}
    \thanks{G. Zhu is with Huawei, Shenzhen, China.}
    \thanks{Corresponding author: Weiqiang Sun (E-mail: sunwq@sjtu.edu.cn).}
}

\markboth{IEEE Journal Paper,~2026}%
{Ye \MakeLowercase{\textit{et al.}}: Beyond Traffic Matrix: DELTA}

\maketitle

\begin{abstract}

The rapid scaling of large language models (LLMs) exacerbates communication bottlenecks in AI data centers (AIDCs). To overcome this, optical circuit switches (OCSs) are increasingly adopted for their superior bandwidth capacity and energy efficiency. However, their reconfiguration overhead precludes intra-iteration topology updates, necessitating a priori engineering of a static topology to absorb time-varying LLM traffic. Existing methods engineer these topologies based on traffic matrices. However, this representation obscures the bursty concurrent bandwidth demands dictated by parallelization strategies and fails to account for the independent channels required for concurrent communication.

To address this, we propose DELTA, an efficient logical topology optimization framework for AIDCs that leverages the computation-communication directed acyclic graph (DAG) to encode time-varying traffic patterns into a mixed-integer linear programming (MILP) model, while exploiting the temporal slack of non-critical tasks to save optical ports without penalizing iteration makespan. By introducing a variable-length time interval formulation, DELTA significantly reduces the solution space compared to the fixed-time-step formulation. To scale to thousand-GPU clusters, we design a dual-track acceleration strategy that combines search space pruning (reducing complexity from quadratic to linear) with heuristic hot-starting. Evaluations on large-scale LLM workloads show that DELTA reduces communication time by up to 17.5\% compared to state-of-the-art traffic-matrix-based baselines. Furthermore, the framework reduces optical port consumption by at least 20\%. Dynamically reallocating these surplus ports to bandwidth-bottlenecked workloads reduces their performance gap relative to ideal non-blocking electrical networks by up to 26.1\%, ultimately enabling most workloads to achieve near-ideal performance.

\end{abstract}


\section{Introduction}
\label{Introduction}

Driven by the vision of Artificial General Intelligence (AGI) and the scaling laws of LLMs \cite{kaplan_scaling_2020a, hoffmann_training_2022}, AIDCs are undergoing rapid expansion. However, the evolution of computational demand has outpaced the upgrade cadence of underlying network equipment, making communication an increasingly prominent bottleneck for cluster scaling \cite{zhang_network_2020, erdil_data_2024b, gholami_ai_2024, benyahya_mosaic_2025, wei_communication_2025}. Existing multi-tier electrical Clos networks that support massive GPU interconnections struggle to bridge this gap. Specifically, as endpoint GPU communication rates surge, relying exclusively on electrical packet switches (EPSs) not only introduces severe bottlenecks in power consumption \cite{qian_alibaba_2024, tian_progress_2024, benyahya_mosaic_2025, bradsmith_new_2025, ashkanseyedi_scaling_2025} and latency \cite{bradsmith_new_2025, ashkanseyedi_scaling_2025, ding_photonic_2025a}, but frequent upgrades to match these endpoint rates also incur prohibitive infrastructure churn \cite{poutievski_jupiter_2022, urata_mission_2022, liu_lightwave_2023}. Consequently, the industry is introducing OCSs--leveraging their low power, low latency, and data-rate transparency--to replace EPSs in the original core or spine layers, driving the emergence of OCS-AIDCs that enable efficient inter-pod optical connectivity while retaining flexible intra-pod electrical switching \cite{poutievski_jupiter_2022, urata_mission_2022, roykim_introducing_2023, wang_topoopt_2023, liu_lightwave_2023, jouppi_tpu_2023, zu_resiliency_2024}.

However, unlike the flexible any-to-any connectivity of EPSs, the coarse-grained, point-to-point connectivity of OCSs struggles to accommodate the dynamic traffic patterns in LLM training. Furthermore, while an OCS can reconfigure its optical circuits within tens of milliseconds \cite{urata_mission_2022, liu_lightwave_2023, liao_mixnet_2025}, the subsequent initialization of transceivers, NICs, and EPSs, including routing-table updates, requires several seconds \cite{zerwas_what_2021, zhang_gemini_2021, liao_mixnet_2025}. Given that mainstream LLM training iterations typically span only a few seconds \cite{nvidia_nemotron4_2024, jiang_megascale_2024, jin_megascalemoe_2025, feng_optimus_2025, liang_lumos_2025}, intra-iteration topology reconfiguration is impractical. Under this constraint, maintaining a static topology throughout an iteration is a pragmatic strategy for current deployments \cite{poutievski_jupiter_2022, liu_lightwave_2023, wang_topoopt_2023, jouppi_tpu_2023, zu_resiliency_2024}. Consequently, a priori engineering of a logical topology (i.e., the number of optical circuits allocated between pod pairs \cite{zhao_understanding_2021,poutievski_jupiter_2022, han_highly_2025}) tailored to absorb time-varying LLM traffic is critical for mitigating OCS-induced communication bottlenecks.

While such a priori engineering is imperative, tailoring a rigid logical topology to accommodate dynamic traffic patterns extends far beyond simple volume-based allocation derived from the conventional traffic-matrix representation. Simply encoding traffic features into an aggregated traffic matrix within an iteration obscures the bursty concurrent communication demands dictated by parallelization strategies, thereby failing to account for the independent channels required for concurrent communication. To capture these transient peaks, existing approaches often employ fine-grained temporal snapshots of traffic. Yet, they typically revert to volume-based allocation within each time slice, aiming to drive dynamic reconfigurations (such as MixNet \cite{liao_mixnet_2025} and other practices in cloud data centers \cite{teh_metteor_2020, zhang_gemini_2021, cao_trod_2021, poutievski_jupiter_2022, teh_enabling_2023})--a path precluded by the aforementioned reconfiguration overhead. Ultimately, solving such an engineering problem hinges on strategically encoding the temporal dynamics of LLM traffic into the optimization formulation of the logical topology.

To address this engineering problem and construct an optimal logical topology, we propose DELTA, a \underline{D}AG-aware, \underline{E}fficient OCS \underline{L}ogical \underline{T}opology optimization  framework for \underline{A}IDCs. We first analyze the spatiotemporal characteristics of LLM training traffic to identify the specific opportunities and optimization challenges in mapping these dynamics to topology construction (Section \ref{section2}). To address these challenges, we introduce a computation-communication DAG of LLM training that dynamically encodes the time-varying traffic patterns into the optimization formulation. This encoding allows the framework to exploit the full spectrum of traffic features while ensuring that the resulting topology targets the exact bottlenecks governing the training iteration time. Based on this DAG, we formulate DELTA-Joint, an MILP model that jointly optimizes the topology and communication scheduling. Specifically, by introducing a variable-length time interval formulation, DELTA-Joint averts the dimensionality curse inherent in conventional fixed-time-step methods. Moreover, by exploiting the temporal slack of non-critical communication tasks, the model incorporates a lexicographic objective to eliminate redundant port allocations without compromising the optimal iteration time (Section \ref{section3}). To scale to large-scale clusters, we design a dual-track acceleration strategy (Section \ref{section4}). Specifically, we employ a search space pruning technique that reduces the problem complexity from quadratic to linear, reducing the MILP solving time to minutes. In parallel, we develop DELTA-Fast, a fast heuristic that yields high-quality topologies and provides a hot start for the MILP, further accelerating MILP solving.
 
Evaluations on large-scale training workloads (e.g., the 671B-parameter DeepSeek model) show that DELTA reduces communication time by up to 17.5\% compared to state-of-the-art traffic-matrix-based baselines. Furthermore, DELTA reduces optical port consumption by at least 20\% without penalizing iteration time. By reallocating these freed ports to bottlenecked workloads, DELTA reduces their performance gap relative to ideal non-blocking electrical networks by up to 26.1\%, enabling most workloads under optical switching to achieve nearly the performance of ideal electrical networks.

In summary, the key contributions of this work are threefold:

\begin{itemize} [leftmargin=*, labelsep=5pt]
    \item We dynamically encode the time-varying traffic patterns of LLM training into the logical topology optimization formulation via a computation-communication DAG. This moves beyond traffic matrices, ensuring we fully exploit traffic characteristics while precisely targeting the bottlenecks that dictate iteration makespan.
    \item We significantly reduce the MILP solution space of DELTA by introducing a variable-length time interval formulation. Combined with search space pruning (reducing complexity from quadratic to linear) and heuristic hot-starting, we solve the MILP within minutes, even at a thousand-GPU scale.
    \item We integrate a resource-saving paradigm into DELTA that exploits temporal slack in non-critical communication tasks to reduce optical port consumption without penalizing iteration time. Dynamically reallocating these freed ports enables most workloads under OCS to achieve performance nearly identical to ideal electrical networks.
\end{itemize}
    
\section{Background and Motivation}
\label{section2}

To motivate the design of DELTA, this section first characterizes the spatiotemporal traffic features of LLM training to identify topology optimization opportunities. We then articulate the algorithmic challenges in mapping these features to OCS configuration, followed by a review of related literature.

\subsection{Features of LLM Training Traffic and Opportunities in Optimizing Logical Topology for OCS-AIDC}
\label{Communication_Characteristics_Section}

In contrast to the stochastic traffic patterns typical of cloud data centers, the communication within LLM training exhibits unique spatiotemporal characteristics driven by parallel strategies, namely Tensor (TP), Pipeline (PP), Data (DP), and Expert (EP) parallelism. Fig.~\ref{LLM_Training_Canvas} profiles a GPT-7B training iteration, revealing three key traffic features and corresponding optimization opportunities for OCS logical topologies:

\begin{figure}[!h]
    \raggedleft 
    \begin{subfigure}[c]{0.476\textwidth}
        \includegraphics[width=\linewidth]{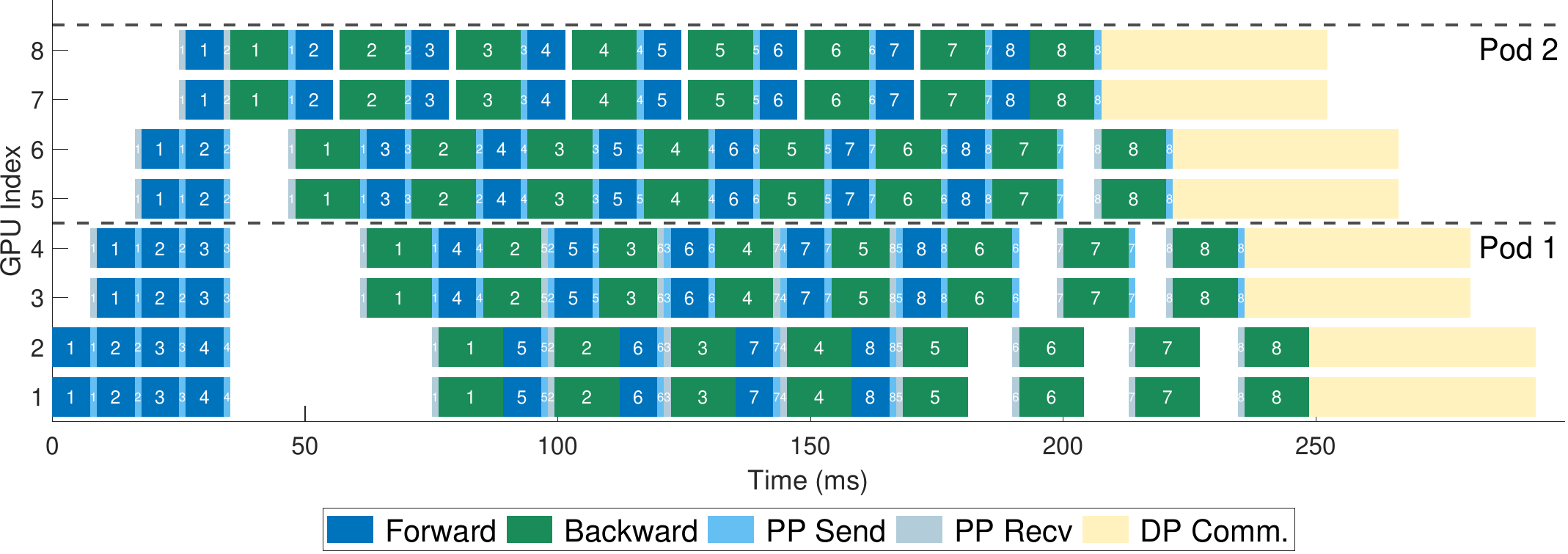}
        \caption{Execution trace of 1F1B scheduling on the first DP replica under a TP=2/PP=4/DP=2 parallelism configuration with 8 micro-batches per GPU (TP communications are subsumed into the forward/backward blocks for visual brevity).}
        \label{llm_timeline}
    \end{subfigure}
    \begin{subfigure}[c]{0.49\textwidth}
        \includegraphics[width=\linewidth]{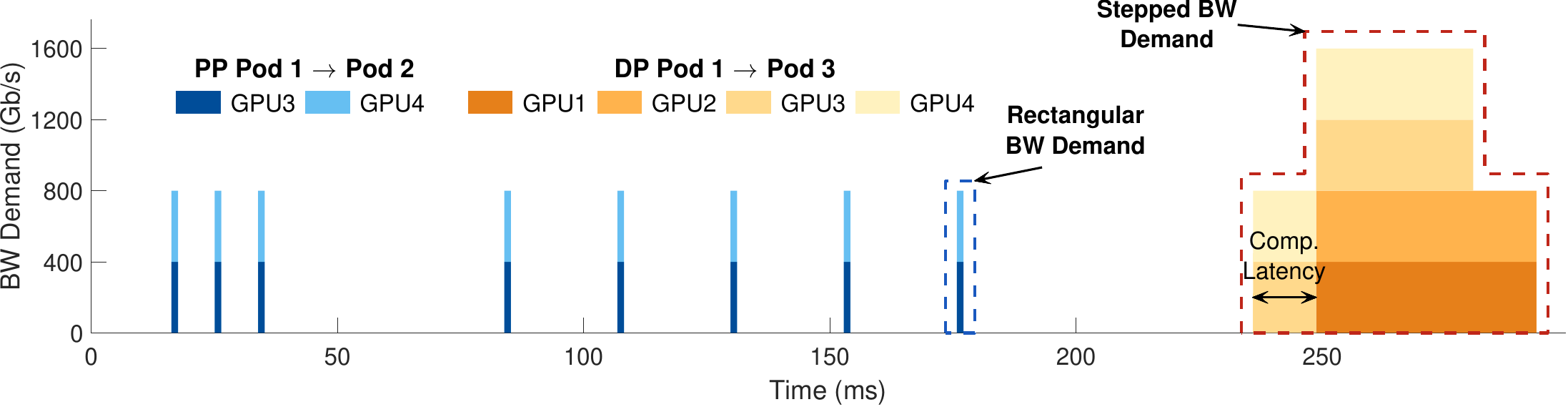}
        \caption{Inter-pod bandwidth demand of pod 1. Note the distinct profiles of PP (rectangular) and DP (stepped) bandwidth demand.}
        \label{cross-pod_traffic}
    \end{subfigure}
    \caption{Spatiotemporal profiling of GPT-7B training schedule and traffic under ideal 400 Gb/s network (GPUs are deployed uniformly across 4 pods).}
    \label{LLM_Training_Canvas}
\end{figure}

\textbf{F1: Communication in LLM training exhibits highly predictable traffic patterns governed by a computation-communication DAG.} Unlike traditional cloud data centers, where traffic is macroscopically predictable but microscopically stochastic \cite{zhang_gemini_2021, poutievski_jupiter_2022}, LLM training exhibits fine-grained predictability. The model architecture and parallelization strategy define a computation-communication DAG that strictly determines the physical properties (e.g., source-destination pairs, data volumes) and causal dependencies of communication. Consequently, as illustrated by the 1F1B execution trace in Fig.~\ref{llm_timeline}, the launch time and scheduling sequence of each task are determined by the chosen parallelization schedule.
\begin{opportunity}
This DAG-encoded predictability in LLM training enables precise topology engineering, allowing network resources to be pre-allocated more precisely to match the bandwidth demands of each communication task.
\end{opportunity}

\textbf{F2: LLM training generates bursty traffic, whose peak demand is driven by the scale of parallelism-induced concurrent flows.} In contrast to the smooth traffic patterns typical of traditional cloud computing, LLM training traffic is characterized by burstiness and concurrency in the temporal dimension \cite{qian_alibaba_2024, gangidi_rdma_2024}. As illustrated in Fig.~\ref{cross-pod_traffic}, inter-pod communication bandwidth demand surges during specific training process phases. Specifically, the concurrent transmission of activations and gradients between model layers (PP communication) from GPU 3 and GPU 4 demands an aggregated inter-pod bandwidth of $2 \times 400$~Gb/s. Parameter synchronization between replicas (DP communication) involves successive communication from GPU 3\&4 and GPU 1\&2, resulting in a maximum bandwidth demand of $4 \times 400$~Gb/s.
\begin{opportunity}
These concurrent bursts imply that to alleviate congestion, resource allocation should prioritize provisioning independent physical channels (i.e., OCS lightpaths) that match flow concurrency, rather than simply allocating resources based solely on aggregated traffic matrices.
\end{opportunity}

\textbf{F3: DP communication exhibits time-varying concurrency and stepped bandwidth demand profiles.} As illustrated in Fig.~\ref{llm_timeline}, in hybrid parallel training, the backward pass propagates in reverse pipeline order, triggering gradient synchronization sequentially from the final stage back to the first. Consequently, the aggregated bandwidth demand manifests as a stepped envelope (Fig.~\ref{cross-pod_traffic})—peaking exclusively during periods of flow overlap—rather than as a uniform rectangular burst typical of PP traffic. This intrinsic temporal staggering facilitates link time-multiplexing. Specifically:
\begin{opportunity}
Sufficiently staggering DP flows across pipeline stages shrinks peak-concurrency intervals, allowing us to provision fewer independent optical channels without exacerbating congestion.
\end{opportunity}

\begin{opportunity}
Furthermore, the temporal slack of earlier DP flows enables dynamic bandwidth reallocation among overlapping DP tasks. Leveraging this slack helps accommodate bandwidth-intensive workloads under stricter capacity constraints without prolonging the global iteration makespan. 
\end{opportunity}

\subsection{Challenges of Optimizing Logical Topology for OCS-AIDC}

\begin{figure}[!h]
    \centering
    \begin{subfigure}[c]{0.24\textwidth}
        \includegraphics[width=\linewidth]{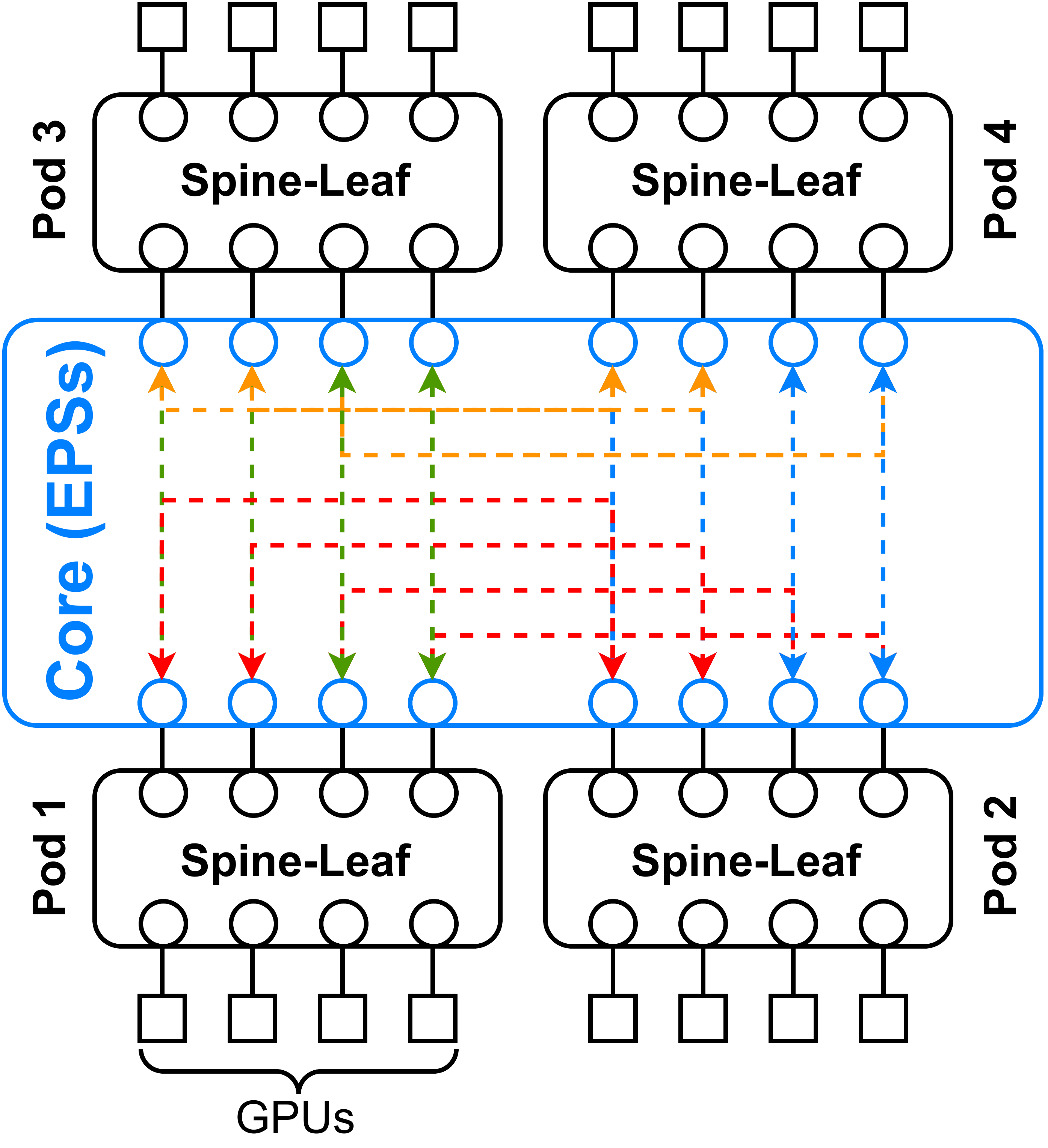}
        \caption{Any-to-any inter-pod paths established by the EPSs.}
        \label{Cross_Pod_EPS_Config}
    \end{subfigure}
        \begin{subfigure}[c]{0.24\textwidth}
        \includegraphics[width=\linewidth]{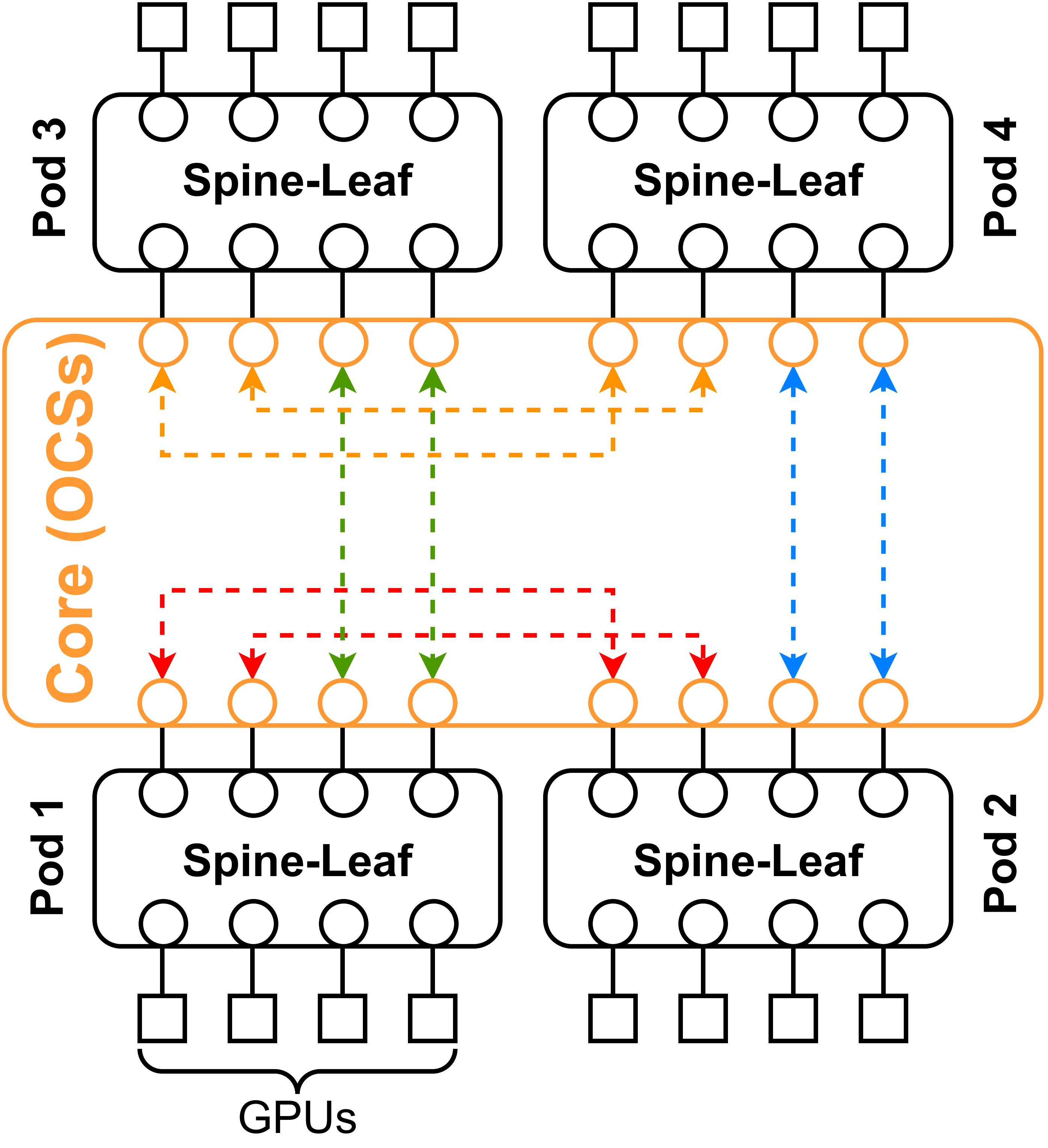}
        \caption{Point-to-point inter-pod paths established by the OCSs.}
        \label{Cross_Pod_OCS_Config}
    \end{subfigure}
    \caption{Comparison of EPS and OCS inter-pod connection patterns for accommodating the LLM training communication demands in Fig.~\ref{LLM_Training_Canvas}. For visual clarity, multiple EPSs/OCSs within the core layer are depicted as a single switching plane.}
    \label{Cross_Pod_OCS_EPS_Config}
\end{figure}

The transition from the any-to-any connectivity of EPSs (Fig.~\ref{Cross_Pod_EPS_Config}) to the rigid point-to-point connectivity of OCSs introduces bottlenecks, as flows with diverging destinations (e.g., orthogonal PP and DP traffic) contend for limited optical exit ports (Fig.~\ref{Cross_Pod_OCS_Config}). Under such hardware constraints, topology construction is essentially a constrained resource allocation problem. While leveraging the aforementioned opportunities is promising, capturing these benefits is nontrivial due to two distinct challenges in encoding the dynamic LLM training traffic patterns into the optimization formulation.

\textbf{C1: Challenge of encoding shifting communication bottlenecks into the optimization objective of reducing training iteration time.} In contrast to cloud data center workloads, where flow completion time linearly correlates with application performance \cite{dukkipati_why_2006, zhang_gemini_2021}, the iteration time of LLM training is determined by the most time-consuming sequence of causally chained computation and communication tasks (hereafter referred to as the critical path). Consequently, accelerating a specific communication flow yields no marginal gain if it resides outside this path. Compounding this issue, these bottlenecks are decision-dependent; optimizing one communication task can shift the critical path to a different sequence of tasks. Thus, a key challenge lies in mathematically encoding these shifting bottlenecks into the topology optimization formulation, ensuring the solver strictly minimizes iteration time for predictable end-to-end performance gains.

\textbf{C2: Challenge of encoding diverse communication demands and formulating an optimization algorithm to resolve their nonlinear coupling with topological decisions.} Encoding the time-varying bandwidth requirements of LLM workloads (\textbf{F2} and \textbf{F3}) into the optimization formulation presents two algorithmic hurdles. First, communication demands driven by different parallelization strategies exhibit distinct temporal behaviors, such as the stepped bandwidth demand envelopes of DP traffic versus the uniform, rectangular demand of PP traffic. Constructing a topology solely based on a traffic matrix fails to capture these temporal dynamics, thereby missing the temporal multiplexing opportunities created by the pipeline-induced staggering of DP flows (i.e., failing to exploit \textbf{O3} and \textbf{O4}). Second, a circular dependency exists between topology decisions and traffic profiles. Specifically, any adjustment to the logical topology inherently changes transmission rates, thereby lengthening or shortening the durations of overlapping flows. This nonlinear coupling renders open-loop static optimization ineffective, necessitating a solver capable of dynamically capturing how topological configurations reshape temporal traffic profiles.

\subsection{Related Work}

\textbf{OCS in cloud data centers.} Existing studies have proposed various topology construction methods for the deployment of OCS in cloud data centers \cite{zhao_minimal_2019a, teh_metteor_2020, zhang_gemini_2021, cao_trod_2021, teh_enabling_2023, hanauer_fast_2022, poutievski_jupiter_2022, hanauer_dynamic_2023, wang_leaf_2024, han_highly_2025}. For instance, \cite{teh_metteor_2020, zhang_gemini_2021, cao_trod_2021, poutievski_jupiter_2022, teh_enabling_2023} construct logical topologies by exploiting long-term traffic pattern prediction to optimize network throughput or link utilization. Meanwhile, works such as \cite{zhao_minimal_2019a, hanauer_fast_2022, hanauer_dynamic_2023, wang_leaf_2024, han_highly_2025} focus on developing algorithms with polynomial time complexity to compute logical topology mapping and reconfiguration schemes that satisfy the physical port mapping constraints of OCS and maximize network throughput. However, these approaches target traditional cloud computing environments featuring stochastic, slow-varying traffic, prioritizing general network performance metrics over job-level execution times. Consequently, they are ill-suited for the deterministic, bursty traffic of LLM training, where minimizing the iteration makespan is the primary objective.

\textbf{Introducing OCS into AI data centers.} Recent studies have begun integrating OCS into AI data centers to optimize AI workload training time while reducing network costs \cite{liu_online_2020, wang_osdl_2022, dong_scheduling_2024, xie_p4incaoi_2025, liu_psscheduler_2025, liao_mixnet_2025, shou_infinitehbd_2025, addanki_when_2025, khani_sipml_2021, wang_topoopt_2023}. Early works \cite{liu_online_2020, wang_osdl_2022} optimized OCS reconfiguration and multiplexed optical circuits by exploiting the interleaved compute-communicate phases of distributed training, but focused exclusively on DP parameter synchronization. To further accelerate this synchronization process, \cite{dong_scheduling_2024, xie_p4incaoi_2025, liu_psscheduler_2025} jointly optimize the schedule of optical circuits and in-network computing resources. However, such DP-centric and hardware-dependent designs lack the generality required for modern LLM training. Other studies \cite{liao_mixnet_2025, shou_infinitehbd_2025, addanki_when_2025} explore architectures where GPUs are directly connected to OCS, reconfiguring the topology based on traffic pattern predictions to optimize the all-to-all or all-reduce communications required for EP or TP/DP. Yet, this approach is incompatible with the prevailing architectural paradigm of scaling clusters by interconnecting multiple pods via OCS. While the methods proposed in \cite{khani_sipml_2021, wang_topoopt_2023, wu_actina_2025} support OCS-based pod-level interconnection and are not restricted to specific workloads, their decision-making processes still rely on aggregated traffic matrices and fail to incorporate the iteration time of LLM training as an optimization objective. Consequently, they do not resolve the previously mentioned challenges \textbf{C1} and \textbf{C2}, which may lead to suboptimal topological decisions--a critical limitation that our work specifically aims to overcome. 

\begin{figure*}[!b]
    \centering
    \begin{subfigure}[c]{0.98\textwidth}
        \includegraphics[width=\linewidth]{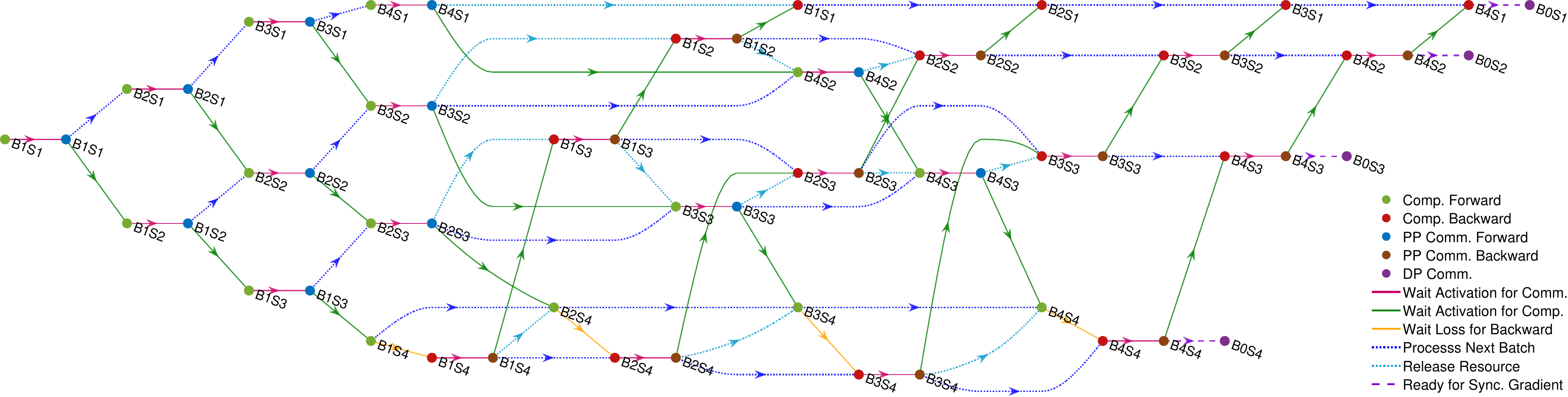}
        \caption{Complete computation-communication DAG of an LLM training iteration.}
        \label{All_Events_Dependency_Graph}
    \end{subfigure}
    
    \begin{subfigure}[c]{0.98\textwidth}
        \medskip
        \includegraphics[width=\linewidth]{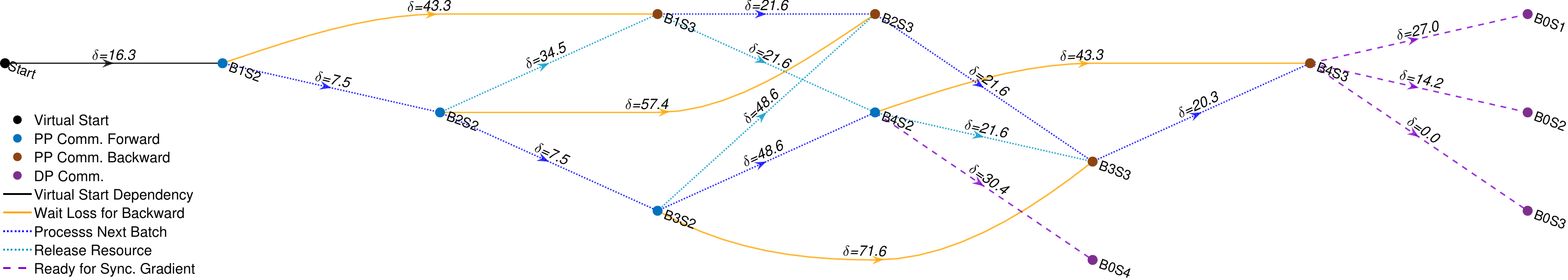}
        \caption{Reduced inter-pod communication DAG, with intra-pod tasks merged into edges weighted by fixed intervals $\delta$ (ms).}
        \label{Dependency_Graph}
    \end{subfigure}
    \caption{Complete computation-communication DAG and reduced inter-pod communication DAG of an LLM training iteration (setup identical to Fig.~\ref{LLM_Training_Canvas}, but with 4 micro-batches). Node label ``B$b$S$s$'' indicates the processing of micro-batch $b$ at stage $s$.}
    \label{scatter}
\end{figure*}

\textbf{Computation-communication DAG-aware communication optimization.} Existing works \cite{hashemi_tictac_2019, jayarajan_prioritybased_2019, peng_generic_2019, bao_preemptive_2020, li_fold3d_2023, li_multidimensional_2024} emphasize the importance of leveraging computation-communication dependencies to optimize communication in LLM training. Specifically, studies such as \cite{hashemi_tictac_2019, jayarajan_prioritybased_2019, peng_generic_2019, bao_preemptive_2020} focus on the parameter synchronization phase in DP, utilizing priority-based scheduling for communication operations according to their dependency order to minimize synchronization stalls. For PP+DP hybrid parallelism, \cite{li_fold3d_2023} groups and reorders the backward computation and corresponding gradient synchronization across different model layers; this removes a portion of DP communication from the critical path of conventional 1F1B scheduling, enabling it to overlap with the computations of other layers. Furthermore, TriRace \cite{li_multidimensional_2024}, an asynchronous pipeline scheduling approach, reschedules the communication required for PP and DP by analyzing the critical path within the DAG, thereby reducing the bubble rate in PP. While the aforementioned works primarily exploit DAG dependencies to schedule communication over a fixed network topology, our approach integrates DAG dependencies into logical topology optimization, tailoring the topology configuration to accelerate LLM training.

\section{Computation-communication DAG-aware Optimization of Logical Topology for OCS-AIDC}
\label{section3}

This section details our approach to optimizing the logical topology for OCS-AIDCs. To resolve the aforementioned challenges, our core strategy is to dynamically encode the time-varying, decision-dependent traffic patterns of LLM training into the optimization space via a computation-communication DAG. Building upon this DAG, we formulate an MILP model structured around variable-length time intervals to efficiently solve this logical topology optimization problem.

\subsection{Encoding LLM Training Traffic Dynamics via the Computation-Communication DAG}

Fig.~\ref{All_Events_Dependency_Graph} illustrates the computation-communication DAG corresponding to the LLM training iteration shown in Fig.~\ref{llm_timeline}. For clarity, the micro-batch size is set to 4, and each pair of synchronously executed PP send and receive tasks is consolidated into a single node. In this DAG, each computation or communication task is represented as a node with a specific execution duration, and the directed edges enforce strict causal dependencies between tasks. Specifically, these dependencies fall into three categories: (1) data dependencies dictated by the computational graph of the LLM itself (\textit{Wait Activation/Loss for Comm./Comp./Backward}); (2) scheduling dependencies governed by the micro-batch scheduling mechanism in PP (\textit{GPU Process Next Batch} and \textit{Release GPU Resource}); and (3) gradient dependencies induced by waiting for the backward pass of the final micro-batch to complete before gradient synchronization (\textit{Ready for Sync. Gradient}). Crucially, once the model architecture and parallelization strategy are defined, this topology-agnostic DAG is determined.

Inspired by \textbf{F1} and \textbf{O1}, we leverage this DAG to dynamically encode the time-varying traffic dynamics. Specifically, we formulate the execution of a training iteration as a precedence-constrained scheduling problem to minimize the overall iteration makespan $C$. This makespan is strictly defined by the length of the critical path:
\begin{equation} \label{All_Events_Dependency_C}
C = \sum_{m \in \mathcal{M}_{\text{crit}}} \tau_m,
\end{equation}
where $\mathcal{M}_{\text{crit}}$ denotes the set of tasks on a critical path of the DAG, and $\tau_m$ represents the duration of individual tasks. Within this formulation, we treat the start and completion times of each task as decision variables bounded by the DAG's edges. The topology decisions dictate the bandwidth, which in turn determines the duration of individual tasks ($\tau_m$). Constrained by the causal dependencies defined by the DAG, any change in a task's duration cascades to its successors, thereby shifting the overall temporal distribution of traffic. By natively encoding this topology-traffic coupling into the decision space via this DAG, our formulation intrinsically captures the decision-dependent communication demands (addressing \textbf{C2}), while simultaneously ensuring that minimizing the objective directly targets the exact communication bottlenecks governing the iteration makespan (addressing \textbf{C1}).

Notably, OCS topology decisions affect only inter-pod communication durations. We therefore reduce the full DAG in Fig.~\ref{All_Events_Dependency_Graph} by eliminating intra-pod computation and communication tasks, whose durations are topology-independent, yielding the inter-pod communication DAG in Fig.~\ref{Dependency_Graph}. Specifically, any chain of intra-pod tasks between two inter-pod communication tasks $m_{\text{pre}}$ and $m$ is replaced by a weighted directed edge $(m_{\text{pre}}, m, \delta_{m_{\text{pre}} \to m})$, where $\delta_{m_{\text{pre}} \to m}$ is the total duration of the eliminated tasks. For intra-pod tasks preceding the first inter-pod communication task, we introduce a virtual inter-pod task at $t=0$ and apply the same replacement. Let $\mathcal{D}$ denote the edge set of this reduced DAG. Accordingly, $C$ can be reformulated as:
\begin{equation} \label{Cross_Pod_Dependency_C}
C = \sum_{m \in \mathcal{M}_{\text{crit}}^{\text{inter-pod}}} \tau_{m}
    + \sum_{(m_{\text{pre}}, m, \delta_{m_{\text{pre}} \to m}) \in \mathcal{D}_{\text{crit}}}
    \delta_{m_{\text{pre}} \to m},
\end{equation}
where $\mathcal{M}_{\text{crit}}^{\text{inter-pod}}$ denotes the inter-pod tasks on the critical path, and $\mathcal{D}_{\text{crit}} \subseteq \mathcal{D}$ denotes the corresponding adjacent dependency edges on the reduced critical path. Eq.~\eqref{Cross_Pod_Dependency_C} provides a reduced-complexity model for LLM training iteration time, which will be implicitly incorporated into the formulation of the MILP in the subsequent section.

\subsection{MILP Formulation with Variable-Length Time Intervals for Logical Topology Optimization}
\label{MILP_Formulation}

Conventional network optimization literature \cite{kandula_calendaring_2014, mohamed_optimizing_2020, li_delayaware_2021, fatemipour_costeffective_2022, wang_linkslice_2022, diaz_nonperiodic_2025, xie_p4incaoi_2025} typically relies on fixed-time-step formulations to capture the bandwidth reallocation and rate variations induced by the dynamic arrival and completion of communication tasks. However, this approach imposes a trade-off between temporal precision and computational efficiency: coarse-grained discretization obscures transient traffic variations, whereas fine-grained discretization inflates the decision space, rendering the model computationally intractable. Empirically, solving the fixed-time-step MILP (detailed in Appendix \ref{app_FTS_MILP}) for small-scale workloads—where training iterations span only hundreds of milliseconds—still requires tens of hours at a 0.1-ms resolution, precluding its real-world application.

\begin{figure}[!h]
    \centering
    \includegraphics[width=\columnwidth]{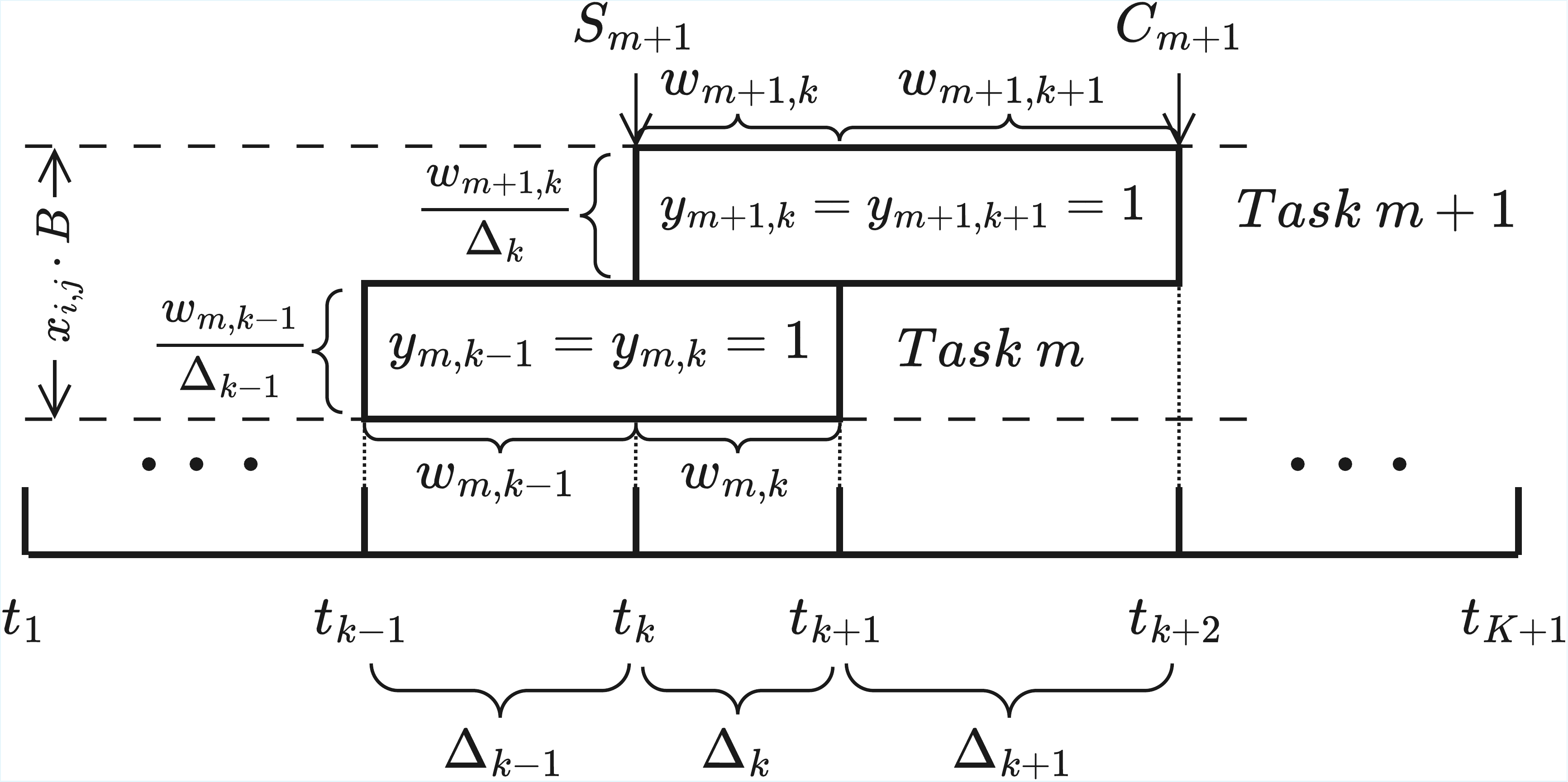}
    \caption{Primary decision variables and their relationships within the variable-length time interval MILP formulation.}
    \label{MILP_Summary}
\end{figure}

To overcome this computational bottleneck, we introduce a variable-length time interval modeling approach inspired by the methodology of discrete event simulation (DES), as illustrated in Fig.~\ref{MILP_Summary}. In a DES paradigm, the system state--specifically, the task activation state ($y_{m,k}$) and flow rates (the ratio of data volume $w_{m,k}$ to interval duration $\Delta_k$)--remains entirely static between state-transition events, which in our context correspond to the initiation ($S_m$) and completion ($C_m$) of communication tasks. Therefore, instead of uniformly slicing the time horizon, we formulate the time interval durations ($\Delta_k$) as decision variables and derive the temporal discretization points ($t_k$) from the occurrences of these events, dynamically partitioning the timeline into a sequence of $K$ variable-length intervals.\footnote{Theoretically, because each task contributes exactly two state-transition timestamps (initiation and completion), at most $2|\mathcal{M}| - 1$ intervals (where $\mathcal{M}$ is the set of all tasks) are required to capture all state transitions. In practice, $K$ is profiled from a baseline simulation and is typically much smaller, as synchronized tasks (e.g., equivalent communication tasks executing concurrently within separate DP replicas in identical network environments) share identical event timestamps.}

Leveraging the stability within each of these $K$ intervals allows a concise set of variables (e.g., $y_{m,k}$, $w_{m,k}$, and $\Delta_k$) to sufficiently characterize the entire temporal horizon, obviating the need to track the system at fixed, fine-grained time steps. This approach fundamentally decouples the problem size from temporal granularity, thereby avoiding the variable explosion inherent in high-resolution time slicing and significantly reducing the MILP solution space. The rigorous mathematical formulation is detailed below.

\vspace{1ex}
\noindent\textbf{Input Parameters:}
\begin{itemize} [leftmargin=*, labelsep=5pt]
    \item $\mathcal{P}$: The set of all pods in the OCS-AIDC.
    \item $\mathcal{G}$: The set of all GPUs in the OCS-AIDC.
    \item $\mathcal{M}$: The set of inter-pod communication tasks involved in a single training iteration. An element $m \in \mathcal{M}$ is defined by a 6-tuple $m \triangleq (i_m, j_m, F_m, V_m, \mathcal{G}^\text{src}_m, \mathcal{G}^\text{dst}_m)$, representing the source pod, destination pod, number of concurrent flows (where communication between a pair of GPUs corresponds to one flow), data volume, source GPU set, and destination GPU set, respectively. Notably, to reduce the problem scale, synchronous communication tasks triggered from multiple GPUs and sharing the same source-destination pod pair (e.g., PP/DP communication tasks from all GPUs within the same stage) are aggregated into a single communication task, with their flow counts accumulated accordingly. Note that \textbf{O2} is exploited through the joint incorporation of $F_m$ and $V_m$ into the formulation, ensuring that resource allocation matches instantaneous flow concurrency rather than relying solely on data volume.
    \item $\mathcal{D}$: The set of temporal dependencies among communication tasks. An element $(m_{\text{pre}}, m, \delta_{m_{\text{pre}} \to m}) \in \mathcal{D}$ indicates that the successor task $m$ can start only after a rigid time interval of $\delta_{m_{\text{pre}} \to m}$ following the completion of the predecessor task $m_{\text{pre}}$. Note that incorporating the a priori knowledge of $\mathcal{M}$ and $\mathcal{D}$ (the DAG of LLM training) exploits \textbf{O1} for proactive topology engineering and \textbf{O3} for DP flow-staggering-aware topology optimization.
    \item $\Phi_\text{src}(g)$, $\Phi_\text{dst}(g)$: The set of communication tasks originating from and destined for GPU $g$.
    \item $U_p$: The maximum number of available OCS ports for the training job in pod $p$, generally constrained to the number of allocated GPUs in that pod for fairness purposes.
    \item $L_{i,j}$: The number of binary bits required to represent the maximum number of optical circuits between pod $i$ and pod $j$, defined as $\lfloor \log_2(\min(U_i, U_j)) \rfloor + 1$.
    \item $B$: The traffic injection rate of a single NIC; correspondingly, each optical circuit is modeled as providing bandwidth $B$, as OCS is rate-transparent.
    \item $K$: The maximum index of time intervals within the considered time horizon.
    \item $M$: A large positive constant.
\end{itemize}

Unless otherwise specified, all indices $i, j$ range over $\mathcal{P}$, $g$ ranges over $\mathcal{G}$, $k$ ranges over $\{1, \dots, K\}$, and $m$ belongs to $\mathcal{M}$. The temporal grid contains $K$ intervals and $K+1$ boundary points, with $t_1=0$ and $t_{K+1}$ denoting the terminal boundary of the considered horizon.

\vspace{1ex}
\noindent\textbf{Decision Variables:}
\begin{itemize} [leftmargin=*, labelsep=5pt]
    \item $x_{i,j}$: An integer variable denoting the number of optical circuits/ports allocated from pod $i$ to pod $j$.
    \item $\beta_{i,j,b}$: A binary auxiliary variable that equals 1 if the $b$-th bit in the binary representation of $x_{i,j}$ is 1, and 0 otherwise.
    \item $t_k$: A continuous variable denoting the $k$-th temporal boundary point ($k=1,\dots,K+1$).
    \item $\Delta_k$: A continuous variable denoting the duration of the $k$-th time interval.
    \item $\rho_{i,j,b,k}$: A continuous auxiliary variable. If $\beta_{i,j,b} = 1$, this variable equals $\Delta_k$; otherwise, it evaluates to 0.
    \item $w_{m,k}$: A continuous variable indicating the data volume transmitted by task $m$ during time interval $k$. Note that by formulating both $w_{m,k}$ and $\Delta_k$ as decision variables, the following MILP inherently exploits \textbf{O4} to optimize the transmission rates of competing flows.
    \item $y_{m,k}$: A binary variable indicating whether task $m$ is active during time interval $k$.
    \item $s\_flag_{m,k}$: A binary auxiliary variable indicating whether task $m$ is initiated at time interval $k$.
    \item $S_m, C_m$: Continuous variables denoting the start time and completion time of task $m$, respectively.
    \item $C$: A continuous variable denoting the iteration time of LLM training.
    \item $u_{i,j,k}$: A continuous auxiliary variable denoting the reference per-flow data transmission volume across the aggregated optical circuits from pod $i$ to pod $j$ during the $k$-th time interval. It is introduced to facilitate the modeling of bandwidth-fairness constraints among all active tasks sharing this inter-pod connection.
\end{itemize}

\vspace{1ex}
\noindent\textbf{Objectives:}

The primary optimization objective is to minimize the iteration time of the training task:
\begin{equation}
\label{Primary_Objective}
    \min \quad C.
\end{equation}

To conserve optical port resources by eliminating redundant allocations on non-critical paths (exploiting \textbf{O4}), an optional lexicographic objective is introduced to minimize the total allocated ports. Letting $C^*$ denote the optimal makespan obtained in Eq.~\eqref{Primary_Objective}, this secondary objective is defined as:
\begin{equation}
    \min \quad \sum_{i \in \mathcal{P}} \sum_{j \in \mathcal{P}, j \neq i} x_{i,j}, \quad \text{s.t.} \quad C \le C^*.
\label{Secondary_Objective}
\end{equation}

\vspace{1ex}
\noindent\textbf{Constraints:}

\textit{1) Topology-Related Constraints:}
\begin{equation}
\begin{cases}
\sum_{j \in \mathcal{P}, j \neq i} x_{i,j} \le U_i, & \forall i. \\
\sum_{i \in \mathcal{P}, i \neq j} x_{i,j} \le U_j, & \forall j.
\end{cases}
\label{TX_RX_Limit}
\end{equation}
Eq.~\eqref{TX_RX_Limit} ensures that the total number of outgoing and incoming logical connections for any pod is constrained by its available transmission ports $U_i$ and receiving ports $U_j$, respectively.
\begin{equation}
x_{i,j} = x_{j,i}, \quad \forall i, j.
\label{Symmetry}
\end{equation}
Eq.~\eqref{Symmetry} ensures that every directed logical link is accompanied by a reciprocal circuit, satisfying the bidirectional connectivity required for practical network operation \cite{han_highly_2025}.
\begin{equation}
x_{i,j} = \sum_{b=0}^{L_{i,j}-1} 2^b \cdot \beta_{i,j,b}, \quad \forall i, j.
\label{Binary_Mapping}
\end{equation}
Eq.~\eqref{Binary_Mapping} expands $x_{i,j}$ into its binary representation to facilitate the subsequent linearization of the bilinear term $x_{i,j} \cdot \Delta_k$. Specifically, we adopt binary rather than unary expansion to reduce the dimension of $b$ from $U_{p}$ down to $\log_2U_{p}$.

\textit{2) Optical Circuit and NIC Capacity-Related Constraints:}
\begin{equation}
\begin{aligned}
& \begin{cases}
\rho_{i,j,b,k} \le M \cdot \beta_{i,j,b}, \\
\rho_{i,j,b,k} \le \Delta_k, \\
\rho_{i,j,b,k} \ge \Delta_k - M \cdot (1 - \beta_{i,j,b}),
\end{cases} \\
& \forall i, j, \forall k, \forall b \in \{0, \dots, L_{i,j}-1\}.
\end{aligned}
\label{Optical_Link_Capacity}
\end{equation}
Eq.~\eqref{Optical_Link_Capacity} employs the Big-M method \cite{wolsey_integer_2020} to linearize the bilinear term $\beta_{i,j,b} \cdot \Delta_k$. Specifically, the formulation constrains $\rho_{i,j,b,k}$ to equal $\Delta_k$ when $\beta_{i,j,b} = 1$, and enforces $\rho_{i,j,b,k} = 0$ when $\beta_{i,j,b} = 0$, thereby ensuring the equivalence $\rho_{i,j,b,k} = \beta_{i,j,b} \cdot \Delta_k$.
\begin{equation}
\sum_{m \in \mathcal{M}_{(i,j)}} w_{m,k} \le B \cdot \sum_{b=0}^{L_{i,j}-1} (2^b \cdot \rho_{i,j,b,k}), \quad \forall k, \forall i,j.
\label{Aggregated_Traffic}
\end{equation}
Eq.~\eqref{Aggregated_Traffic} bounds the aggregated traffic rate in $\Delta_k$ by the total bandwidth of the allocated optical circuits from pod $i$ to pod $j$. Here, $\mathcal{M}_{(i,j)}$ denotes the set of tasks originating from pod $i$ and destined for pod $j$. Note that the term $\sum_{b=0}^{L_{i,j}-1} (2^b \cdot \rho_{i,j,b,k})$ is introduced to represent the binary expansion and linearization of the bilinear term $x_{i,j} \cdot \Delta_k$.
\begin{equation}
\begin{cases}
\sum_{m \in \Phi_\text{src}(g)} \frac{w_{m,k}}{F_m} \le B \cdot \Delta_k, & \forall k, \forall g. \\
\sum_{m \in \Phi_\text{dst}(g)} \frac{w_{m,k}}{F_m} \le B \cdot \Delta_k, & \forall k, \forall g.
\end{cases}
\label{Endpoint_Injection}
\end{equation}
Eq.~\eqref{Endpoint_Injection} ensures that the aggregated injection and reception data volume of each GPU during interval $k$ must not exceed the maximum data theoretically transferable by the NIC bandwidth $B$. This constraint assumes a one-to-one mapping between GPUs and NICs, regulating both the outgoing tasks in $\Phi_\text{src}(g)$ and the incoming tasks in $\Phi_\text{dst}(g)$.

\textit{3) Data Conservation and Task Activation Constraints:}
\begin{equation}
\sum_{k=1}^{K} w_{m,k} = V_m, \quad \forall m.
\label{Data_Conservation}
\end{equation}
Eq.~\eqref{Data_Conservation} ensures that the sum of the data transmitted by a task $m$ across all time intervals exactly equals its predefined total data volume, $V_m$.
\begin{equation}
w_{m,k} \le V_{m} \cdot y_{m,k}, \quad \forall m, k.
\label{Inactive_Cutoff}
\end{equation}
Eq.~\eqref{Inactive_Cutoff} restricts a task to transmit data strictly during the time intervals in which it is active (i.e., $y_{m,k}=1$). If a task is inactive ($y_{m,k}=0$), its transmission volume $w_{m,k}$ is constrained to 0.
\begin{equation}
\begin{cases}
s\_flag_{m,k} \ge y_{m,k} - y_{m,k-1}, & \forall m, k \quad (\text{setting } y_{m,0} = 0). \\
\sum_{k=1}^{K} s\_flag_{m,k} = 1, & \forall m.
\end{cases}
\label{Global_Continuity}
\end{equation}
Eq.~\eqref{Global_Continuity} introduces $s\_flag_{m,k}$ to identify the activation event (rising edge) of task $m$. By limiting the total number of such events to one, the constraint prevents task suspension, ensuring that each task occupies a single, contiguous time block.

\textit{4) Temporal Boundary and Mapping Constraints:}
\begin{equation}
\begin{cases}
\Delta_k = t_{k+1} - t_k, & \forall k. \\
t_1 = 0.
\end{cases}
\label{Time_Continuity}
\end{equation}
Eq.~\eqref{Time_Continuity} defines the duration of the $k$-th time interval, $\Delta_k$, as the difference between adjacent temporal points.
\begin{equation}
\begin{cases}
S_m \le t_k + M \cdot (1 - y_{m,k}), & \forall m, k. \\
C_m \ge t_{k+1} - M \cdot (1 - y_{m,k}), & \forall m, k.
\end{cases}
\label{Start_Completion_Time}
\end{equation}
Eq.~\eqref{Start_Completion_Time} defines the temporal boundaries of task $m$ based on its active intervals. By utilizing the Big-M method, it ensures that the interval $[S_m, C_m]$ encompasses all time slots $\Delta_k$ where the task is scheduled to transmit ($y_{m,k}=1$).

\textit{5) Inter-pod Communication DAG Constraints:}
\begin{equation}
S_{m} \ge C_{m_{\text{pre}}} + \delta_{m_{\text{pre}} \to m}, \quad \forall (m_{\text{pre}}, m, \delta_{m_{\text{pre}} \to m}) \in \mathcal{D}.
\label{DAG_Dependency}
\end{equation}
Eq.~\eqref{DAG_Dependency} ensures that the execution order strictly adheres to the reduced inter-pod communication DAG. Specifically, it ensures that any successor task $m$ can be initiated only after its predecessor $m_{\text{pre}}$ has completed and an additional interval $\delta_{m_{\text{pre}} \to m}$ has elapsed.

\textit{6) Flow Fairness Constraints (Optional for Comparison):}
\begin{equation}
\label{Flow_Fairness}
\begin{aligned}
& \begin{cases}
\frac{w_{m,k}}{F_m} - u_{i,j,k} \le M \cdot (1 - y_{m,k}), \\
u_{i,j,k} - \frac{w_{m,k}}{F_m} \le M \cdot (1 - y_{m,k}),
\end{cases} \\
& \forall i, j, \forall m \in \mathcal{M}_{(i,j)}, \forall k.
\end{aligned}
\end{equation}
Eq.~\eqref{Flow_Fairness} is an optional constraint that simulates conventional fair-sharing mechanisms by compelling concurrently active tasks ($y_{m,k}=1$) traversing the same inter-pod connection to transmit an equal volume of data per flow ($u_{i,j,k}$). It is introduced as a comparative baseline to quantify the performance gains of the optimized rate control strategy against the conventional fair-sharing policy.

\textit{7) Objective Function Constraint:}
\begin{equation}
C \ge C_m, \quad \forall m.
\label{Objective_Function}
\end{equation}
Eq.~\eqref{Objective_Function} defines the LLM training iteration time $C$ as the upper bound of all individual task completion times $C_m$. Note that, rather than explicitly modeling $C$ using the critical-path-based formulation in Eq.~\eqref{Cross_Pod_Dependency_C}, Eq.~\eqref{Objective_Function} relies on the linear temporal constraints derived from the DAG $\mathcal{D}$ (Eq.~\eqref{DAG_Dependency}) to express $C$ in an MILP-tractable format.

\section{Dual-Track Acceleration by MILP Search Space Pruning and DES-Accelerated Heuristics}
\label{section4}

Despite the substantial simplification achieved by the reduced inter-pod communication DAG and the variable-length time-interval formulation, directly solving the resulting MILP remains computationally prohibitive for thousand-GPU-scale multi-pod LLM training. To overcome this, this section presents a dual-track acceleration framework that combines MILP search-space pruning with a DES-accelerated heuristic. First, we leverage domain-specific strategies to reduce the size of the MILP. In parallel, we develop a DES-accelerated heuristic search algorithm to rapidly explore topologies and provide hot-start solutions to the MILP solver. Together, these two tracks enable the optimizer to obtain high-quality solutions within minutes, even for thousand-GPU deployments.

\subsection{Strategies for Pruning MILP Search Space}
\label{acc_MILP}

The scale of the MILP is dominated by the task-time decision variables ($w_{m,k}$, $y_{m,k}$, and $s\_flag_{m,k}$). Because the number of intervals $K$ grows linearly with the number of tasks $|\mathcal{M}|$, the number of these variables increases quadratically, i.e., $\mathcal{O}(|\mathcal{M}|^2)$. By contrast, the number of topology and capacity variables ($x_{i,j}$, $\beta_{i,j,b}$, $\rho_{i,j,b,k}$, and $u_{i,j,k}$) remains relatively limited. Although the number of candidate pod pairs $(i, j)$ is theoretically upper-bounded by $|\mathcal{P}|^2$, the sparsity of communication patterns in LLM training substantially reduces the number of spatial variables that need to be instantiated. Moreover, because the binary expansion index $b$ is logarithmically bounded, the total number of these variables grows at most linearly with the interval count $K$, namely $\mathcal{O}(|\mathcal{M}|)$. As a result, they account for only a negligible portion of the overall model size. Therefore, our acceleration strategies specifically focus on mitigating the $\mathcal{O}(|\mathcal{M}|^2)$ variable growth caused by the task-time decision variables.

\textit{1) Reducing Total Task Count by Isomorphism and Independence of DP Replicas:}

We first project the multi-replica optimization problem onto a single-replica domain by leveraging the isomorphism and independence of DP replicas to mitigate the $\mathcal{O}(|\mathcal{M}|^2)$ complexity associated with task-time variables. Specifically, when DP replicas are deployed in identical network environments, their communication tasks are theoretically synchronized. By formulating the MILP for a single reference replica and mapping the resulting topology configurations and schedules across the others, the global task count $|\mathcal{M}|$ can be reduced to the scale of a single replica. Conversely, when DP replicas span heterogeneous networks, we exploit the decoupled nature of their execution. Aside from parameter synchronization tasks, the operations within each replica are independent of the others. This allows the global optimization to be decomposed into parallel single-replica sub-problems. To satisfy the DP synchronization barrier, the schedule of the bottleneck replica--which initiates the synchronization phase last--is established as the global temporal baseline to align the DP communication across all replicas. Under these assumptions, this methodology preserves solution validity and remains consistent with global iteration-time optimality while reducing the dimensionality of task-time decision variables from an $\mathcal{O}(|\mathcal{M}|^2)$ complexity to a single-replica $\mathcal{O}(|\mathcal{M}_{\text{single-replica}}|^2)$ scale.

\textit{2) Reducing Search Space by Task-Time Domain Pruning:}

\begin{figure}[!h]
    \centering
    \begin{subfigure}[c]{0.24\textwidth}
        \includegraphics[width=\linewidth]{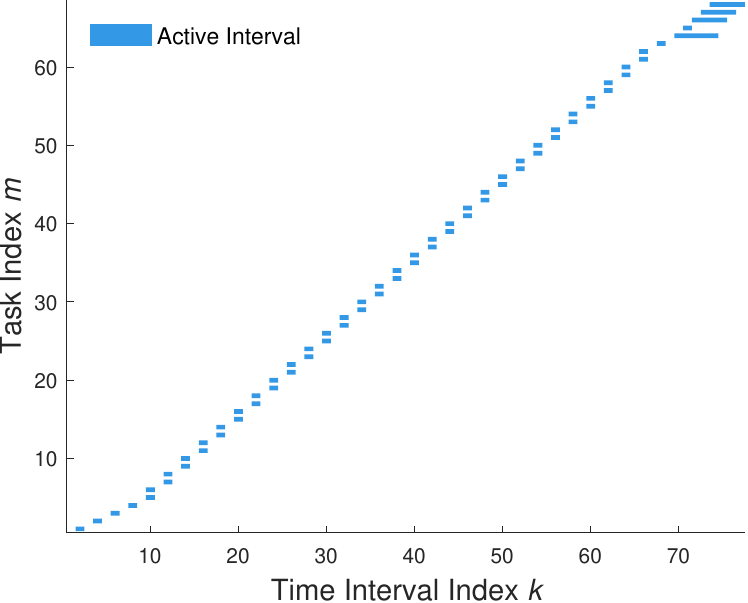}
        \caption{Active intervals of tasks under a baseline simulation.}
        \label{reference_interval}
    \end{subfigure}
    \begin{subfigure}[c]{0.24\textwidth}
        \includegraphics[width=\linewidth]{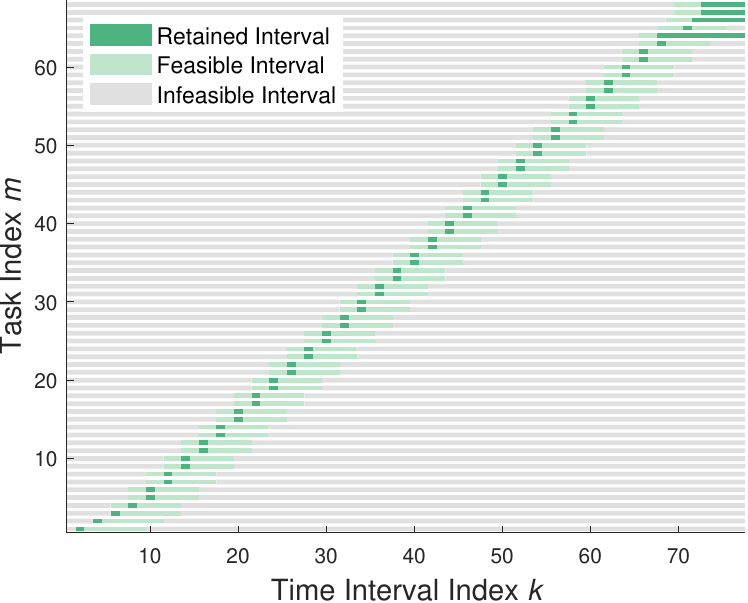}
        \caption{Retained intervals refined by DAG and anchor-guided pruning.}
        \label{retained_interval}
    \end{subfigure}
    \caption{Search space reduction for MILP via index pruning (LLM training setup as in Fig.~\ref{LLM_Training_Canvas}, but with 32 micro-batches).}
    \label{task_interval_window}
\end{figure}

To further reduce computational complexity, we prune the search space of task-time-related decision variables (i.e., $w_{m,k}$, $y_{m,k}$, and $s\_flag_{m,k}$). As illustrated by Fig.~\ref{reference_interval}, within the model and parallel configuration defined in Fig.~\ref{llm_timeline}, each inter-pod communication task for each DP replica occupies only a small fraction of the total time intervals in practice.\footnote{It should be noted that because some inter-pod communication tasks overlap in time (as shown in Fig.~\ref{llm_timeline}, e.g., the forward PP communications for micro-batches $\{4, 5, 6, 7, 8\}$ on Stage 2 and Stage 3 completely overlap with the backward PP communications for micro-batches $\{2, 3, 4, 5, 6\}$), the total number of time intervals $K$ is less than $2|\mathcal{M}_{\text{single-replica}}| - 1$.} Given these observations, it is reasonable to infer that the optimal solution of this MILP will exhibit a sparse structure for $y_{m,k}$ and its associated variables $w_{m,k}$, $s\_flag_{m,k}$.

Specifically, the temporal dependencies defined by the set $\mathcal{D}$ impose strict boundaries on the feasible active intervals for communication tasks. Dictated by these dependencies, a task cannot be activated too early, as this would inevitably force its predecessors into invalid time intervals (i.e., interval indices $< 1$). Conversely, given a total interval count $K$, a task cannot remain active too late without pushing its subsequent tasks into invalid time intervals (i.e., indices $> K$) (provided the assigned $K$ accommodates the theoretical optimal solution, pruning these states preserves optimality). Leveraging this property, we apply a topological sorting algorithm to $\mathcal{D}$—computing the longest paths from the global source and sink nodes of the DAG to each task—to determine the earliest and latest allowable interval indices for every task. Consequently, the corresponding decision variables ($w_{m,k}$, $y_{m,k}$, and $s\_flag_{m,k}$) can be strictly constrained to 0 outside this feasible temporal window, thereby substantially reducing the search space. The pruning effect of this strategy on the $m-k$ search space is illustrated by the light green regions in Fig.~\ref{retained_interval}.

While the preceding topological sorting establishes boundaries based on graph-wide longest paths, relying solely on it yields overly conservative search spaces (detailed in Appendix \ref{app_limitations}). From a local, task-centric perspective, as observed from Fig.~\ref{Dependency_Graph} and Fig.~\ref{reference_interval}, the dependencies in $\mathcal{D}$ strictly constrain the execution of intermediate tasks: a task $m$ cannot overlap with the state transitions of its predecessors or successors. Driven by this strict causal ordering, an intermediate task typically monopolizes a whole time interval, leaving minimal room for shifts within the discrete interval sequence and rendering its active index range highly deterministic. Consequently, a lightweight baseline simulation (Fig.~\ref{reference_interval}) can approximate this range. Exploiting this predictability, we extract index bounds $[\tilde{k}^{\text{start}}_m, \tilde{k}^{\text{end}}_m]$ from the baseline simulation to act as reference anchors that localize the execution of each intermediate task $m$. By integrating these anchoring bounds with the longest-path-derived boundaries, we tightly confine the allowable active indices, further pruning the $m-k$ search space as illustrated by the dark green regions in Fig.~\ref{retained_interval}. The complete algorithm workflow is presented in Alg.~\ref{alg_task_time_pruning}. Since most tasks are intermediate tasks (Fig.~\ref{Dependency_Graph}), bounding their indices significantly tightens the entire MILP scale. Through a cascade effect, this tightening further prunes the indices of DP tasks. Ultimately, the combined restriction on both intermediate and DP tasks reduces the scale of task-time decision variables from $\mathcal{O}(|\mathcal{M}|^2)$ to $\mathcal{O}(|\mathcal{M}|)$.

Collectively, by applying the aforementioned task reduction and search space pruning strategies, our approach empirically reaches high-quality MILP solutions within minutes, even at the thousand-GPU scale.

\begin{algorithm}[htbp]
\caption{TaskTimeIndexPruning}
\label{alg_task_time_pruning}
\begin{algorithmic}[1]
\Require $\mathcal{M}$, $\mathcal{D}$, $K$, anchors $\tilde{k}^{\text{start}}_m$ and $\tilde{k}^{\text{end}}_m$. Let $\mathcal{M}_{\text{succ}} \triangleq \{m \in \mathcal{M} \mid \exists (m, v, \delta) \in \mathcal{D}\}$ denote the subset of tasks that have successor tasks.
\Ensure Pruned upper bounds $UB(\cdot)$ for decision variables.

\vspace{0.5em}
\Statex \hspace{-\algorithmicindent} \textbf{$\triangleright$ Step 1: Initialization \& Anchoring}
\State \textbf{for each} $m \in \mathcal{M}$ \textbf{do} $k^{\min}_m \gets 1, \ k^{\max}_m \gets K$
\State \textbf{for each} $m \in \mathcal{M}_{\text{succ}}$ \textbf{do} $k^{\min}_m \gets \tilde{k}^{\text{start}}_m, \ k^{\max}_m \gets \tilde{k}^{\text{end}}_m$
\State Compute in-degree $deg_{\text{in}}(m)$ and out-degree $deg_{\text{out}}(m)$ based on $\mathcal{D}$.

\vspace{0.5em}
\Statex \hspace{-\algorithmicindent} \textbf{$\triangleright$ Step 2: Forward Bounds Propagation (Earliest Start)}
\State $\mathcal{Q} \gets \{m \in \mathcal{M} \mid deg_{\text{in}}(m) = 0\}$
\While{$\mathcal{Q} \neq \emptyset$}
    \State $u \gets \mathcal{Q}.\text{pop}()$
    \For{\textbf{each} $(u, v, \delta) \in \mathcal{D}$}
        \State $k^{\min}_v \gets \max\big(k^{\min}_v, k^{\min}_u + 1 + \mathbb{I}[\delta > 0]\big)$
        \State $deg_{\text{in}}(v) \gets deg_{\text{in}}(v) - 1$; \textbf{if} $deg_{\text{in}}(v) == 0$ \textbf{then} $\mathcal{Q}.\text{push}(v)$
    \EndFor
\EndWhile

\vspace{0.5em}
\Statex \hspace{-\algorithmicindent} \textbf{$\triangleright$ Step 3: Backward Bounds Propagation (Latest End)}
\State $\mathcal{Q} \gets \{m \in \mathcal{M} \mid deg_{\text{out}}(m) = 0\}$
\While{$\mathcal{Q} \neq \emptyset$}
    \State $v \gets \mathcal{Q}.\text{pop}()$
    \For{\textbf{each} $(u, v, \delta) \in \mathcal{D}$}
        \State $k^{\max}_u \gets \min\big(k^{\max}_u, k^{\max}_v - 1 - \mathbb{I}[\delta > 0]\big)$
        \State $deg_{\text{out}}(u) \gets deg_{\text{out}}(u) - 1$; \textbf{if} $deg_{\text{out}}(u) == 0$ \textbf{then} $\mathcal{Q}.\text{push}(u)$
    \EndFor
\EndWhile

\vspace{0.5em}
\Statex \hspace{-\algorithmicindent} \textbf{$\triangleright$ Step 4: Variable Domain Pruning}
\For{\textbf{each} $m \in \mathcal{M}$}
    \State $\mathcal{K}_{\text{invalid}} \gets \{1, \dots, k^{\min}_m - 1\} \cup \{k^{\max}_m + 1, \dots, K\}$
    \State \textbf{for each} $k \in \mathcal{K}_{\text{invalid}}$ \textbf{do} $UB(y_{m,k}), UB(w_{m,k}), UB(s\_flag_{m,k}) \gets 0$
\EndFor
\State \textbf{return} $UB(\cdot)$

\end{algorithmic}
\end{algorithm}

\subsection{DES-Accelerated Heuristics for Fast Topology Search}
\label{sim_base_GA}

Although the pruned MILP formulation presented in Section \ref{acc_MILP} enables efficient joint topology and flow-rate optimization, broader scenarios, such as GPU-resource allocation co-optimization in multi-tenant clusters, demand frequent invocations of the topology optimizer. To this end, we propose a fast, DES-accelerated heuristic search algorithm (DELTA-Fast). The key idea is to decouple the search space by offloading the constraint-solving process of the cumbersome task-time dimensions (i.e., $y_{m,k}$, $w_{m,k}$, and $\Delta_k$) to a lightweight DES engine, thereby avoiding computationally expensive joint searches across all variables. Specifically, an outer-loop genetic algorithm optimizes the logical topology ($x_{i,j}$), while an inner-loop DES simulates task execution over time based on DAG dependencies. Through a single simulation pass, it determines valid values for these temporal and task-state variables and implicitly satisfies all constraints. Although this decoupling sacrifices the dynamic joint optimization of flow rates (degenerating to a fair-sharing mechanism), it effectively reduces the computational burden, enabling the rapid generation of high-quality topology candidates for upper-level co-optimization frameworks. Furthermore, since the DES execution trace is isomorphic to the MILP's event-driven formulation, the simulated results seamlessly map to an initial feasible solution, providing an efficient hot start for the MILP. To enable efficient search, we design DELTA-Fast around two core components: a DAG-aware search space pruning strategy and a domain-adapted genetic algorithm.

\textit{1) DAG-Aware Search Space Pruning:}

To prune the search space, we eliminate redundancy by jointly exploiting physical and logical constraints. Physically, NIC-bound GPU data injection limits render any optical circuits ($x_{i,j}$) exceeding the maximum concurrent inter-pod flows redundant (exploiting \textbf{O2}). Logically, the DAG $\mathcal{D}$ prohibits temporally dependent inter-pod tasks from executing concurrently. Motivated by these observations, we propose Alg.~\ref{alg_capacity_bound} to estimate the maximum number of concurrent communication pairs per inter-pod connection and establish an upper bound for $x_{i,j}$: First, similar to Alg.~\ref{alg_task_time_pruning}, we use $\mathcal{D}$ and a coarse estimated iteration time upper bound $\hat{T}_{up}$ to derive the earliest start time ($\text{EST}_{m}$) and latest completion time ($\text{LCT}_{m}$) for each task $m$ (Line 2, detailed in Alg.~\ref{alg_cal_task_time_windows} in Appendix \ref{app_search_space_pruning}). Next, computing the transitive closure $\mathcal{R}$ of $\mathcal{D}$ identifies all dependency-linked, mutually exclusive task pairs (Line 3). Subsequently, for each inter-pod connection $(u,v)$, we merge and sort the temporal boundaries ($\text{EST}$ and $\text{LCT}$) of all associated tasks into a discrete interval sequence $T$ (Lines 5--6). Scanning each interval in $T$, the algorithm extracts active tasks $\mathcal{A}$ to construct a conflict graph $G$, where tasks act as vertices, flow counts as vertex weights, and mutual exclusivity relations as edges (Lines 8--12). Since mutually exclusive tasks cannot run in parallel, the maximum concurrent flow per interval equals the maximum weight independent set (MWIS) of $G$. Finally, the maximum MWIS across all intervals establishes the upper bound $\bar{X}_{u,v}$ (Lines 13--14) of $x_{u,v}$, which constrains the search space for the heuristic algorithm.

\begin{algorithm}[htbp]
\caption{XUpperBoundEstimation}
\label{alg_capacity_bound}
\begin{algorithmic}[1]
\Require $\mathcal{M}$, $\mathcal{D}$, $B$, estimated iteration time upper bound $\hat{T}_{up}$.
\Ensure Upper bound matrix $\bar{X}$ for $x_{i,j}$.

\vspace{0.5em}
\Statex \hspace{-\algorithmicindent} \textbf{$\triangleright$ Step 1: Initialization \& Preprocessing}
\State Initialize $\bar{X}$ as a zero matrix for all pod pairs
\State $[\text{EST}, \text{LCT}] \gets \text{CalTaskTimeWindows}(\mathcal{M}, \mathcal{D}, B, \hat{T}_{up})$
\State $\mathcal{R} \gets \text{TransitiveClosure}(\mathcal{D})$ \Comment{via matrix squaring}

\vspace{0.5em}
\Statex \hspace{-\algorithmicindent} \textbf{$\triangleright$ Step 2: Capacity Bound Computation}
\For{\textbf{each} inter-pod connection $(u, v)$}
    \State $\mathcal{M}_{u,v} \gets \{m \in \mathcal{M} \mid \text{task } m \text{ traverses } (u,v)\}$
    \State $T \gets \text{Sort}\big(\text{Unique}(\{\text{EST}_m, \text{LCT}_m \mid m \in \mathcal{M}_{u,v}\})\big)$

    \For{$k \gets 1$ \textbf{to} $|T|-1$}
        \State $t_{\text{mid}} \gets (T[k] + T[k+1]) / 2$
        \State $\mathcal{A} \gets \{m \in \mathcal{M}_{u,v} \mid \text{EST}_m \le t_{\text{mid}} < \text{LCT}_m\}$

        \If{$\mathcal{A} \neq \emptyset$}
            \State \textbf{$\triangleright$ Construct the conflict graph for MWIS}
            \State $G \gets \text{Graph}(\text{Vertices}=\mathcal{A}, \text{Weights}=F_{m \in \mathcal{A}}, \text{Edges}=\mathcal{R}|_{\mathcal{A}})$

            \State $c_{\text{max}} \gets \text{SolveMWIS}(G)$ \Comment{Solve MWIS}
            \State $\bar{X}_{u,v} \gets \max(\bar{X}_{u,v}, c_{\text{max}})$
        \EndIf
    \EndFor
\EndFor
\State \textbf{return} $\bar{X}$

\end{algorithmic}
\end{algorithm}

\textit{2) DES-Accelerated Domain-Adapted Genetic Algorithm:}

To efficiently navigate the search space and accelerate convergence, we develop a domain-adapted genetic algorithm that integrates a fast DES engine driven by the reduced DAG $\mathcal{D}$ for rapid iteration time evaluation, alongside a topology repair mechanism that uses the capacity upper bounds $\bar{X}$ to restore the physical feasibility of invalid offspring. As illustrated in Alg.~\ref{alg_domain_adapted_genetic}, after initializing a population of feasible topologies strictly bounded by physical port limits $U$ and capacity bounds $\bar{X}$ (Line 1, detailed in Alg.~\ref{alg_feasible_random_init} in Appendix \ref{app_ga_subroutines}), iteration time (fitness) is evaluated in parallel via the DAG-aware DES (Line 2), where the total allocated optical circuits can serve as a secondary fitness. Notably, this DES evaluation can also be accelerated by exploiting the isomorphism and independence of DP replicas to reduce the scale of the simulated task set. Following standard selection, crossover, and mutation operations (Lines 7--8), the $\text{RepairTopo}$ function (Alg.~\ref{alg_repair_topo} in Appendix \ref{app_ga_subroutines}) calibrates the child's optical circuits to restore compliance with $U$ and $\bar{X}$ (Line 9). Unrepairable children are replaced by newly generated feasible configurations to maintain population diversity (Lines 10--11). The remaining steps follow a standard genetic algorithm framework to update the population and track the globally optimal logical topology $X^*$. Collectively, Alg.~\ref{alg_domain_adapted_genetic} ensures efficient convergence toward optimal logical topologies within tens of seconds.

\begin{algorithm}[htbp]
\caption{SimBasedDomainAdaptedGA}
\label{alg_domain_adapted_genetic}
\begin{algorithmic}[1]
\Require Active pod pairs $E$, port capacities $U=[U_1,...,U_{|\mathcal{P}|}]$, upper bounds $\bar{X}$, GA parameters ($N_{\text{pop}}$, $N_{\text{gen}}$), $\mathcal{M}$, $\mathcal{D}$
\Ensure Logical topology $X^*$

\State $\mathcal{X} \gets \left\{ \text{FeasibleRandomInit}(E, U, \bar{X}) \right\}_{i=1}^{N_{\text{pop}}}$
\State $\text{Fitness} \gets \text{ParallelEvalDES}(\mathcal{X}, \mathcal{M}, \mathcal{D})$
\State $X^*, C_{\max}^* \gets \text{GetBestIndividual}(\mathcal{X}, \text{Fitness})$

\For{$gen = 1 \dots N_{\text{gen}}$}
    \State $\mathcal{X}_{\text{new}} \gets \text{RetainElites}(\mathcal{X}, \text{Fitness})$
    \While{$|\mathcal{X}_{\text{new}}| < N_{\text{pop}}$}
        \State $p_1, p_2 \gets \text{TournamentSelection}(\mathcal{X}, \text{Fitness})$
        \State $child \gets \text{Mutate}(\text{Crossover}(p_1, p_2))$
        \State $child, success \gets \text{RepairTopo}(child, E, U, \bar{X})$
        \State \textbf{if not} $success$ \textbf{then} $child \gets \text{FeasibleRandomInit}(E, U, \bar{X})$
        \State $\mathcal{X}_{\text{new}} \gets \mathcal{X}_{\text{new}} \cup \{child\}$
    \EndWhile

    \State $\mathcal{X} \gets \mathcal{X}_{\text{new}}$; $\text{Fitness} \gets \text{ParallelEvalDES}(\mathcal{X}, \mathcal{M}, \mathcal{D})$

    \If{$\min(\text{Fitness}) < C_{\max}^*$}
        \State $X^*, C_{\max}^* \gets \text{GetBestIndividual}(\mathcal{X}, \text{Fitness})$
    \EndIf
\EndFor
\State \Return $X^*$

\end{algorithmic}
\end{algorithm}

\section{Evaluations}
\label{section5}

In this section, we evaluate the performance of the proposed DAG-aware efficient logical topology optimization algorithm and compare it to existing solutions. All evaluations are conducted on a commodity computing platform equipped with an Intel i5-13500H (2.6 GHz) CPU and 32 GB of RAM.

\subsection{Evaluation Setup}

\subsubsection{Workloads}

\begin{table*}[!h]
    \centering
    \caption{Evaluation setup for the four representative LLMs.}
    \label{tab_llm_workloads}
    \begin{tabular}{lccccccc}
        \toprule
        \textbf{Model Name} & \textbf{TP} & \textbf{PP} & \textbf{ETP} & \textbf{EP} & \textbf{\# of GPUs} & \textbf{\# of GPUs / pod / replica} & \textbf{\# of micro-batches} \\
        \midrule
        Megatron-177B & 8 & 6 & - & - & 384 & 16 & 48 \\
        Mixtral-8X22B & 2 & 8 & 1 & 8 & 128 & 16 & 64 \\
        Megatron-462B & 8 & 16 & - & - & 1024 & 32 & 128 \\
        DeepSeek-671B & 2 & 16 & 1 & 8 & 256 & 32 & 128 \\
        \bottomrule
    \end{tabular}
\end{table*}

We evaluate the logical topology optimization problem across four representative LLM training workloads (Megatron-177B, Mixtral-8X22B (MoE model), Megatron-462B, and DeepSeek-671B (MoE model)), with the evaluation configurations provided in Table \ref{tab_llm_workloads}. The parallel strategy configurations for these models are derived from the Megatron benchmarks \cite{liu_moe_2025, nvidiacorporation_performance_2025, nvidiacorporation_megatronlm_2026}. To emulate practical scheduling strategies that increase the number of micro-batches to mitigate pipeline bubbles, the number of micro-batches processed per GPU per iteration is set to 8 times the PP size. The communication demand traces are generated via SimAI \cite{li_aicb_2025, wang_simai_2025}. Furthermore, to simulate the fragmented model deployment typical of multi-tenant clusters, we limit the number of GPUs per pod to 16 per DP replica for the two smaller models and 32 for the two larger models. Notably, allocating more GPUs per pod per replica (i.e., \# of GPUs / pod / replica) would localize more communication tasks within a single pod, inadvertently simplifying the optimization problem. Therefore, we restrict this capacity to ensure sufficient inter-pod tasks, thereby rigorously evaluating the proposed algorithm's performance. Moreover, we restrict inter-pod communication to PP and DP, as standard practice confines the latency-sensitive and bandwidth-intensive EP and TP to intra-pod electrical networks. Unless otherwise specified, the sequence length for LLM training is set to 4096, and inter-pod bandwidth is set to 400 Gb/s/GPU. For fairness, we assume that the maximum number of OCS ports ($U_p$) exclusively allocated to a training job in each pod is strictly bounded by the number of GPUs assigned to that job within the pod.

\subsubsection{Algorithms for Comparison}

Our evaluation includes six optimization algorithms: DELTA-Joint, DELTA-Topo, DELTA-Fast, and three traffic-matrix-based baselines: Prop-Alloc, Sqrt-Alloc, and Iter-Halve, which are implemented following the allocation principles of SiP-ML \cite{khani_sipml_2021}, ACTINA \cite{wu_actina_2025}, and TopoOpt \cite{wang_topoopt_2023}, respectively. Specifically, DELTA-Joint and DELTA-Topo construct the logical topology by solving the MILP formulated in Section \ref{MILP_Formulation}, with the former additionally co-optimizing the flow rates. DELTA-Fast is the heuristic algorithm detailed in Alg.~\ref{alg_domain_adapted_genetic}. Among the baselines, Prop-Alloc treats all traffic as concurrent; under this assumption, minimizing the maximum transmission time corresponds to assigning logical links to communication pairs in proportion to their traffic volumes. Sqrt-Alloc assumes that traffic from the same pod to different destinations is transmitted sequentially (analogous to DP and PP communication within a stage), and thus assigns links proportional to the square root of traffic volume to minimize total transmission time. Iter-Halve repeatedly assigns a logical link to the communication pair with the highest weight, initially set by the traffic matrix, and subsequently halves that weight for the next round of allocation. Unless otherwise specified, all algorithms are run with a time limit of 600 seconds. All MILP instances are solved using Gurobi 13.0. To enhance computational efficiency, the proposed DELTA suite employs four-thread parallelism.

\subsubsection{Performance Metric}

We adopt Normalized Communication Time (NCT) as the primary evaluation metric, defined as the ratio of the inter-pod communication time on the critical path under the OCS architecture to that under an ideal non-blocking electrical network with zero packet-processing delay. This normalization provides a standardized baseline for evaluating algorithms across diverse model architectures. Furthermore, focusing on the critical path reflects the communication bottleneck that governs the actual iteration makespan.

\begin{figure*}[!b] 
    \centering
    \begin{subfigure}[c]{0.24\textwidth} 
        \includegraphics[width=\linewidth]{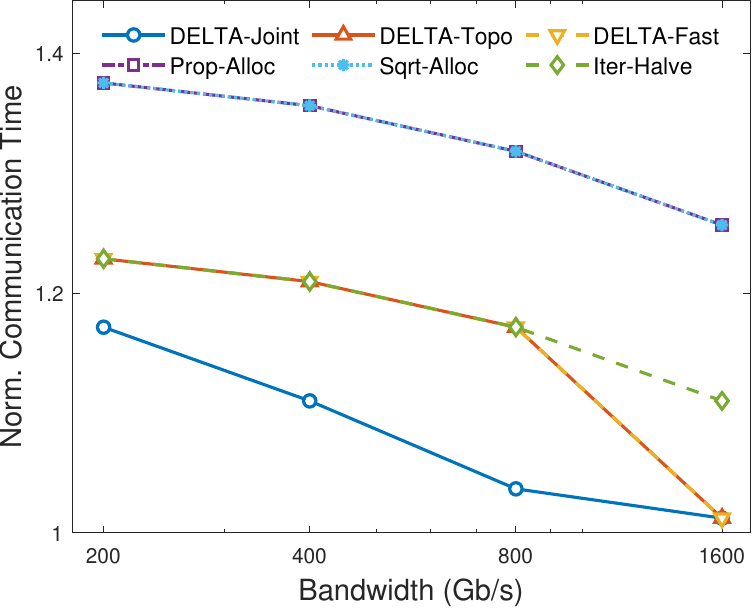}
        \caption{Megatron-177B.}
        \label{Megatron-177B}
    \end{subfigure}\hfill 
    \begin{subfigure}[c]{0.24\textwidth}
        \includegraphics[width=\linewidth]{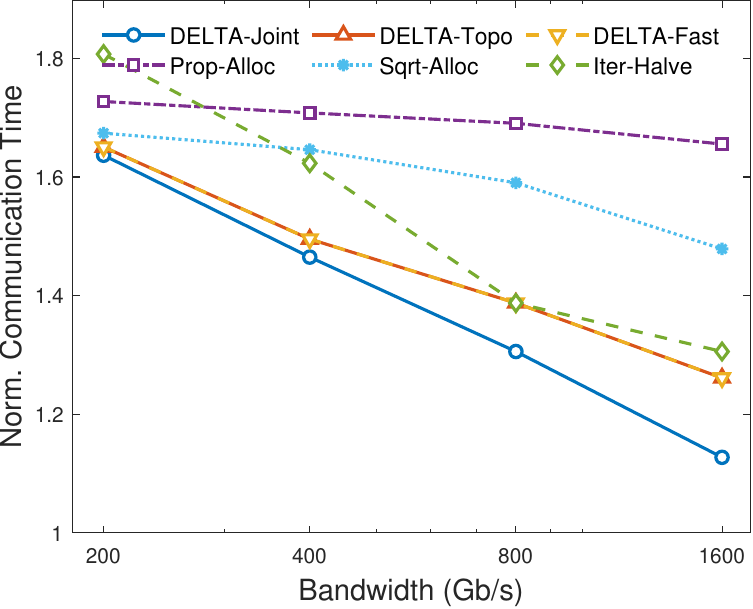}
        \caption{Mixtral-8X22B.}
        \label{Mixtral-8X22B}
    \end{subfigure}\hfill
    \begin{subfigure}[c]{0.24\textwidth}
        \includegraphics[width=\linewidth]{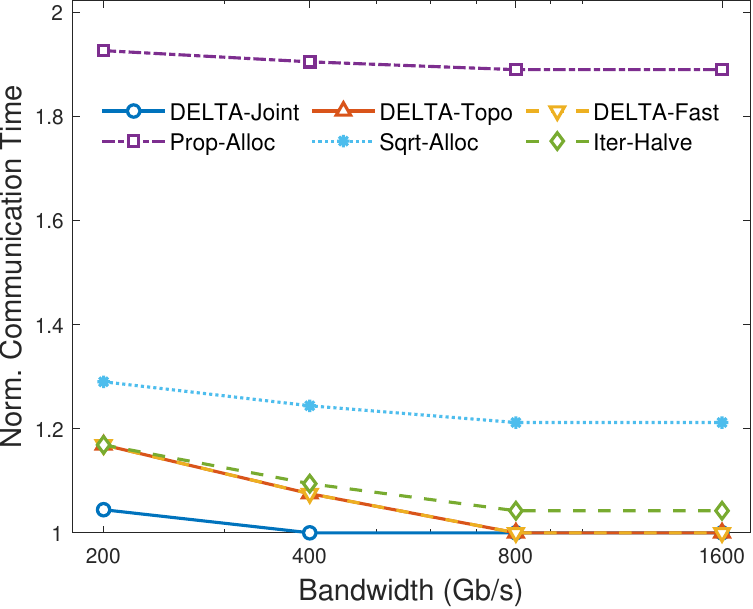}
        \caption{Megatron-462B.}
        \label{Megatron-462B}
    \end{subfigure}\hfill
    \begin{subfigure}[c]{0.24\textwidth}
        \includegraphics[width=\linewidth]{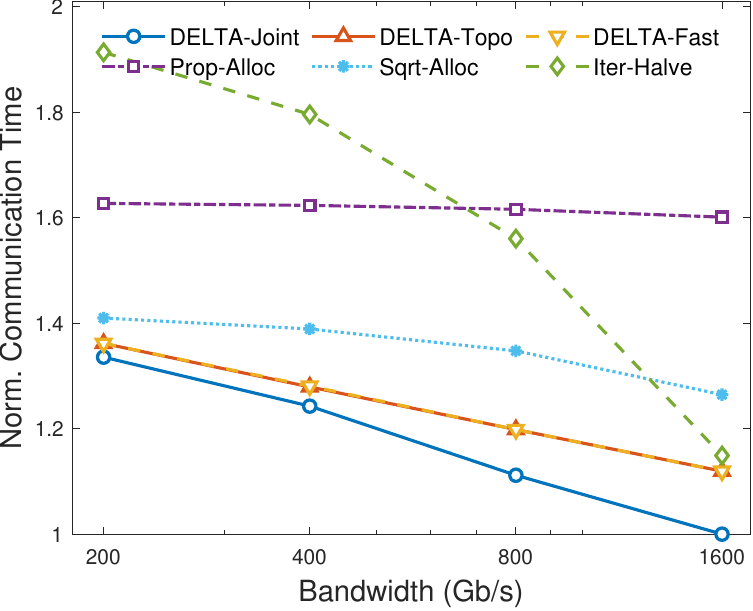}
        \caption{DeepSeek-671B.}
        \label{Deepseek-671B}
    \end{subfigure} 
    \caption{Performance of DAG-driven vs. traffic-matrix-driven topology optimization under varying inter-pod bandwidths.}
    \label{solver_cmp}
\end{figure*}

\subsection{Performance Comparison on Varying Inter-pod Bandwidths}

Fig.~\ref{solver_cmp} illustrates the NCT of the six evaluated algorithms under varying inter-pod bandwidth capacities. As inter-pod bandwidth increases, the communication-to-computation time ratio drops, shortening the relative duration of concurrent inter-pod communication for DP. This mitigates the communication degradation caused by allocating fewer logical links than the maximum number of concurrent flows, thereby reducing the NCT across all algorithms. While all algorithms benefit from this trend, the proposed DAG-driven DELTA suite consistently outperforms traffic-matrix-based baselines. Specifically, across the four evaluated LLMs, DELTA-Joint achieves maximum NCT reductions of 11.5\%, 13.7\%, 10.7\%, and 17.5\% (at bandwidths of 800, 1600, 200, and 800 Gb/s, respectively) compared to the best-performing baselines. Generally, in next-generation network infrastructures with higher inter-pod bandwidths, the NCT reductions achieved by DELTA over the baselines become more substantial.

Within the DELTA suite, DELTA-Joint achieves the lowest NCT, confirming that topology and flow-rate co-optimization can further reduce the time on the critical path compared to topology-only methods. Across the four LLMs, it yields maximum NCT reductions of 11.5\%, 10.6\%, 10.7\%, and 10.7\% (at 800, 1600, 200, and 1600 Gb/s, respectively) over the topology-only DELTA. Moreover, DELTA-Fast performs identically to DELTA-Topo across all evaluated scenarios, validating the efficacy of the heuristic DELTA-Fast.

Notably, DELTA-Joint enables a static optical topology to achieve performance close to that of an ideal, non-blocking electrical switch (NCT = 1), without incurring the overhead of reconfigurations. Specifically, at an inter-pod bandwidth of 1600 Gb/s, DELTA-Joint reduces the performance gap to the theoretical ideal to 1.2\%, 12.7\%, 0\%, and 0\% across the four evaluated LLMs, respectively, demonstrating the power of DAG-aware joint optimization. We also note that at bandwidths of 400 Gb/s or below, communication performance under a static topology remains limited for workloads such as Megatron-177B, Mixtral-8X22B, and DeepSeek-671B. However, this bottleneck can be effectively mitigated by reallocating surplus optical ports saved from other workloads (without compromising their original NCTs) to these bandwidth-bottlenecked workloads, as demonstrated in Section \ref{section_compress_port}.

\begin{figure}[!h]
    \centering
    \includegraphics[width=0.95\columnwidth]{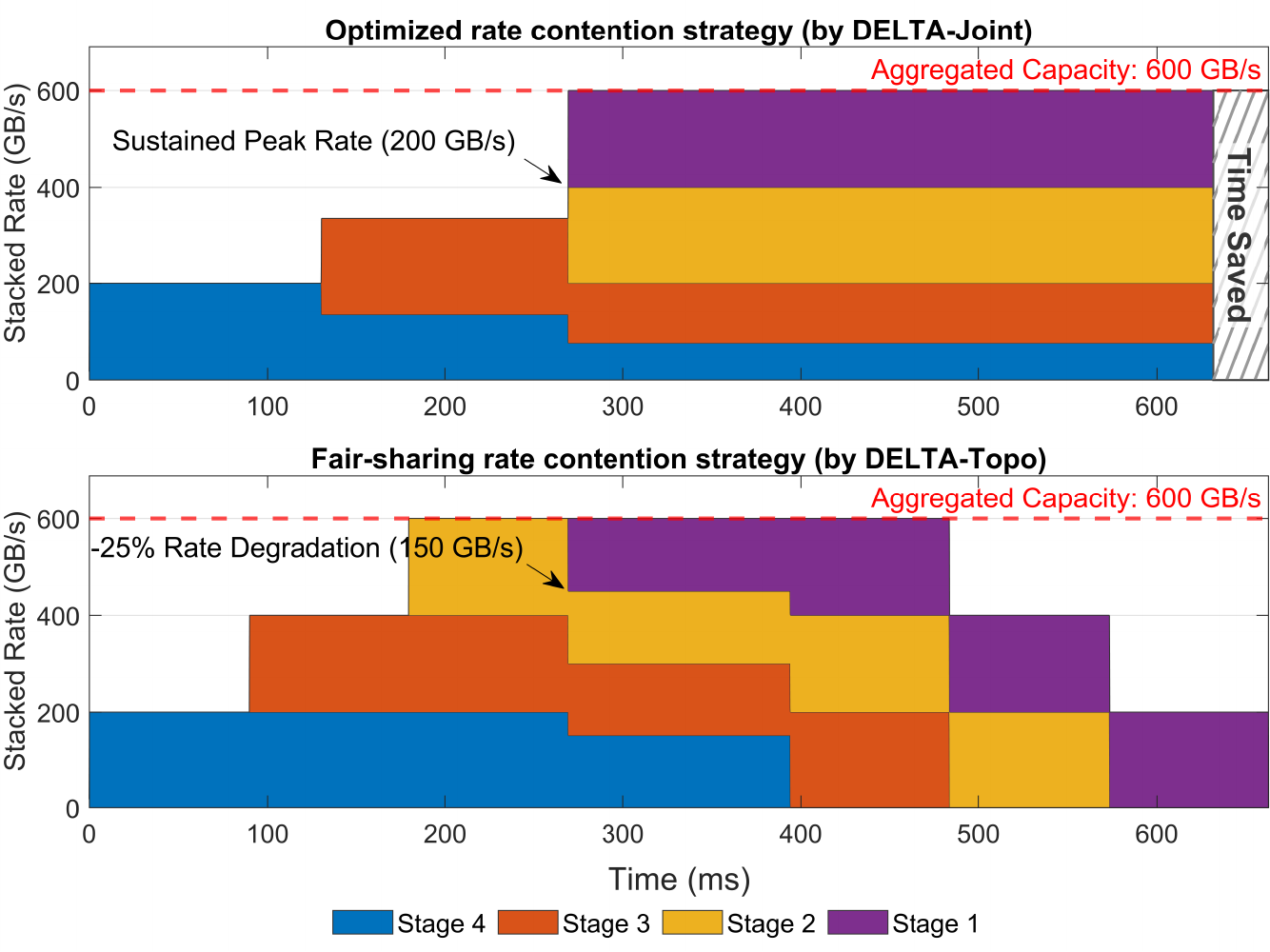}
    \caption{Flow-rate control results of DP communication for Megatron-462B (400 Gb/s). For clarity, the time axis is zeroed at the moment Stage 4 initiates DP communication.}
    \label{Link_Comparison}
\end{figure}

Fig.~\ref{Link_Comparison} illustrates how DELTA-Joint approaches the performance of an ideal non-blocking electrical switch through its optimized flow-rate control strategy. As shown in the figure, when bandwidth contention is induced by concurrent communication from multiple stages, DELTA-Joint ensures that the critical flow (DP communication from Stage 1), which dictates the global iteration makespan, continuously transmits at its physical upper bound of 200 GB/s (i.e., the aggregated injection bandwidth of 4 NICs). In contrast, the conventional fair-sharing strategy employed by DELTA-Topo causes this critical flow to experience a 25\% rate degradation (from 200 to 150 GB/s) during contention periods.

\vspace{-2pt}

This sustained peak transmission in DELTA-Joint is achieved by exploiting the temporal slack of non-critical tasks. During contention periods, DELTA-Joint actively compresses the transmission rates of earlier-initiated flows (Stages 2 to 4) that possess temporal slack and dynamically reallocates the released bandwidth to Stage 1. Consequently, this flow-control strategy effectively prevents local congestion from delaying the critical task.

\begin{figure*}[!b]
    \centering
    \begin{subfigure}[c]{0.24\textwidth} 
        \includegraphics[width=\linewidth]{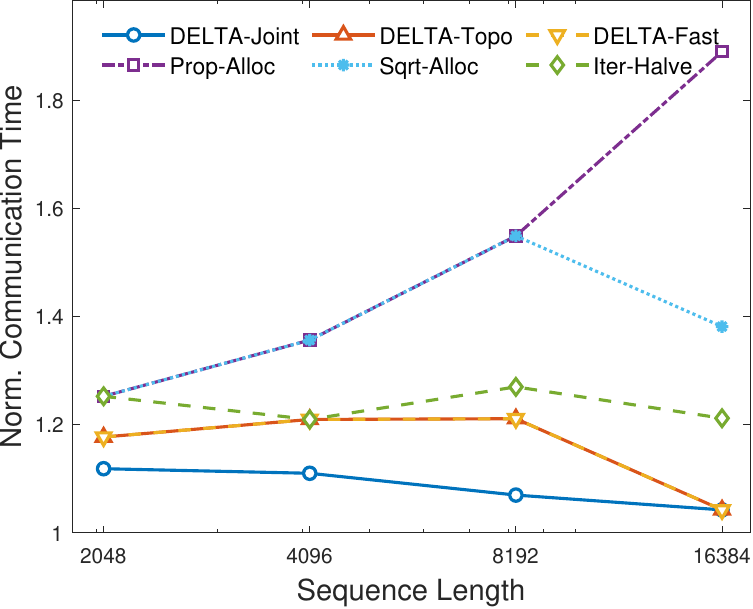}
        \caption{Megatron-177B.}
        \label{Megatron-177B_seq}
    \end{subfigure}\hfill 
    \begin{subfigure}[c]{0.24\textwidth}
        \includegraphics[width=\linewidth]{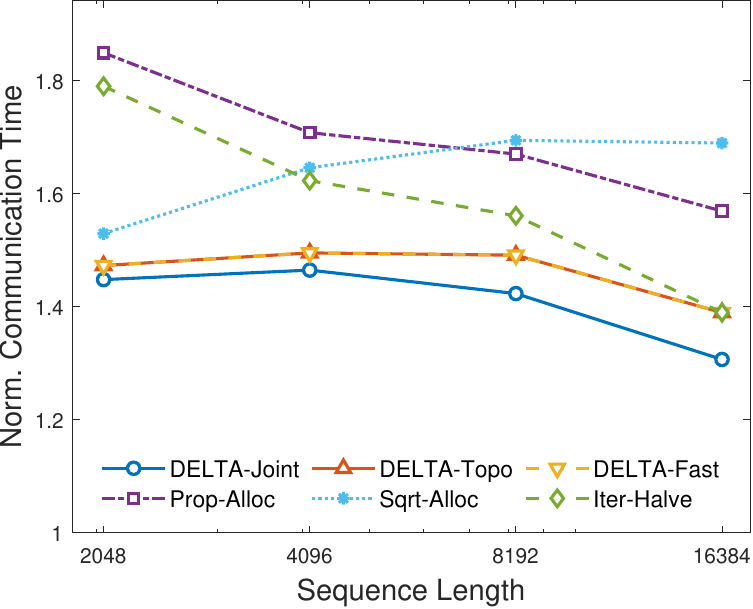}
        \caption{Mixtral-8X22B.}
        \label{Mixtral-8X22B_seq}
    \end{subfigure}\hfill
    \begin{subfigure}[c]{0.24\textwidth}
        \includegraphics[width=\linewidth]{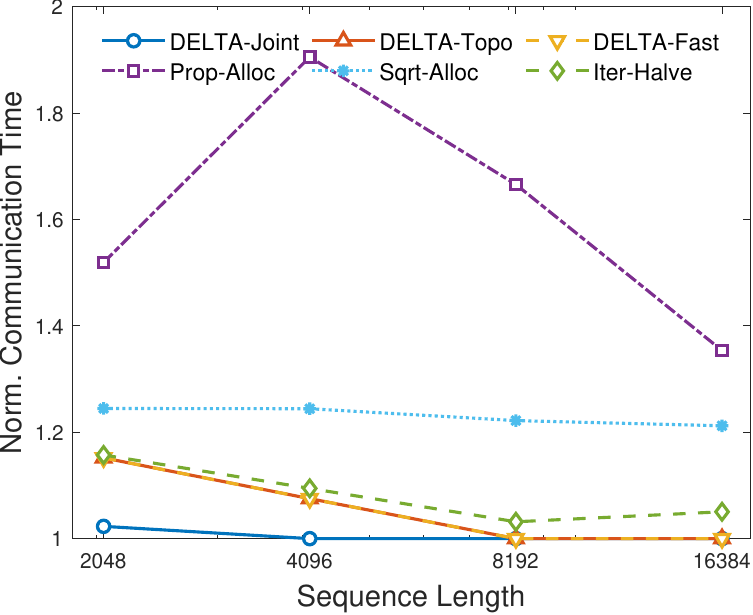}
        \caption{Megatron-462B.}
        \label{Megatron-462B_seq}
    \end{subfigure}\hfill
    \begin{subfigure}[c]{0.24\textwidth}
        \includegraphics[width=\linewidth]{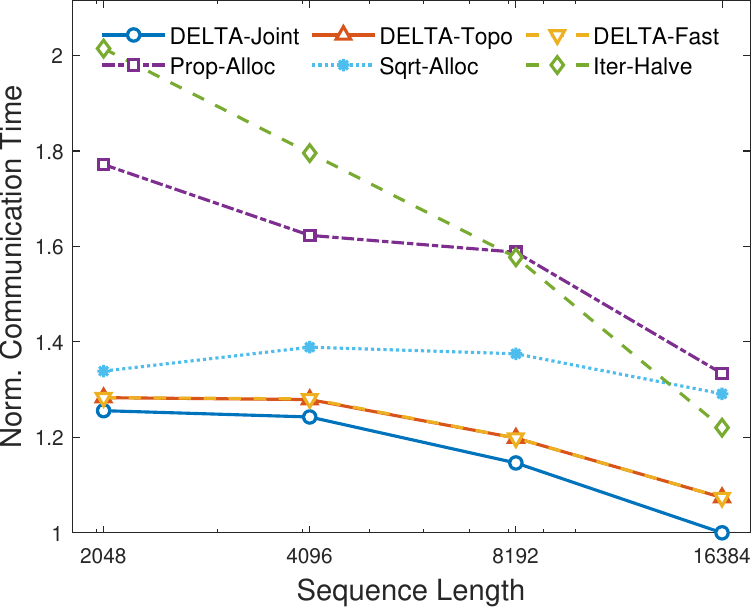}
        \caption{DeepSeek-671B.}
        \label{Deepseek-671B_seq}
    \end{subfigure} 
    \caption{Performance of DAG-driven vs. traffic-matrix-driven topology optimization under varying sequence lengths.}
    \label{solver_cmp_seq}
\end{figure*}

\subsection{Performance Comparison on Varying Sequence Lengths}

Beyond bandwidth, variations in model hyperparameters (e.g., sequence length) also reshape the communication bandwidth demands during LLM training. Consequently, we further investigate the performance of topology optimization algorithms at varying sequence lengths, as shown in Fig.~\ref{solver_cmp_seq}.

As the sequence length increases, the DELTA suite not only consistently outperforms the baseline algorithms but also exhibits a more consistent downward trend in NCT. Specifically, across the four evaluated LLMs, DELTA-Joint achieves maximum NCT reductions of 15.8\%, 9.8\%, 11.6\%, and 18.1\% (at sequence lengths of 8192, 4096, 2048, and 16384, respectively) compared to the best-performing baselines. Moreover, consistent with previous evaluations, DELTA-Fast retains its performance parity with DELTA-Topo. Fundamentally, an increase in sequence length triggers a twofold shift in the communication characteristics of LLM training: on the one hand, prolonged computation intervals (caused by the expanded sequence length) facilitate enhanced staggering of DP traffic, thereby creating better link time-multiplexing opportunities and mitigating concurrency bottlenecks; on the other hand, the concurrent ``rectangular'' bandwidth demands of PP scale proportionally with the sequence length. Leveraging the DAG-aware design, the DELTA suite more effectively navigates the dynamics of these superposed effects.

Furthermore, when evaluated against an ideal non-blocking electrical network, DELTA-Joint exhibits a performance gap of 11.9\%, 44.8\%, 2.3\%, and 25.6\% across the four models at a sequence length of 2048. As the sequence scales to 16384, this gap narrows to 4.2\%, 30.7\%, 0\%, and 0\%, respectively. As the context window of large language models continues to expand, this performance disparity is projected to become increasingly marginal.

\subsection{Reduction of Optical Port Consumption and Performance Gains via Port Reallocation}
\label{section_compress_port}

\begin{figure}[!h]
    \centering
    \includegraphics[width=0.65\columnwidth]{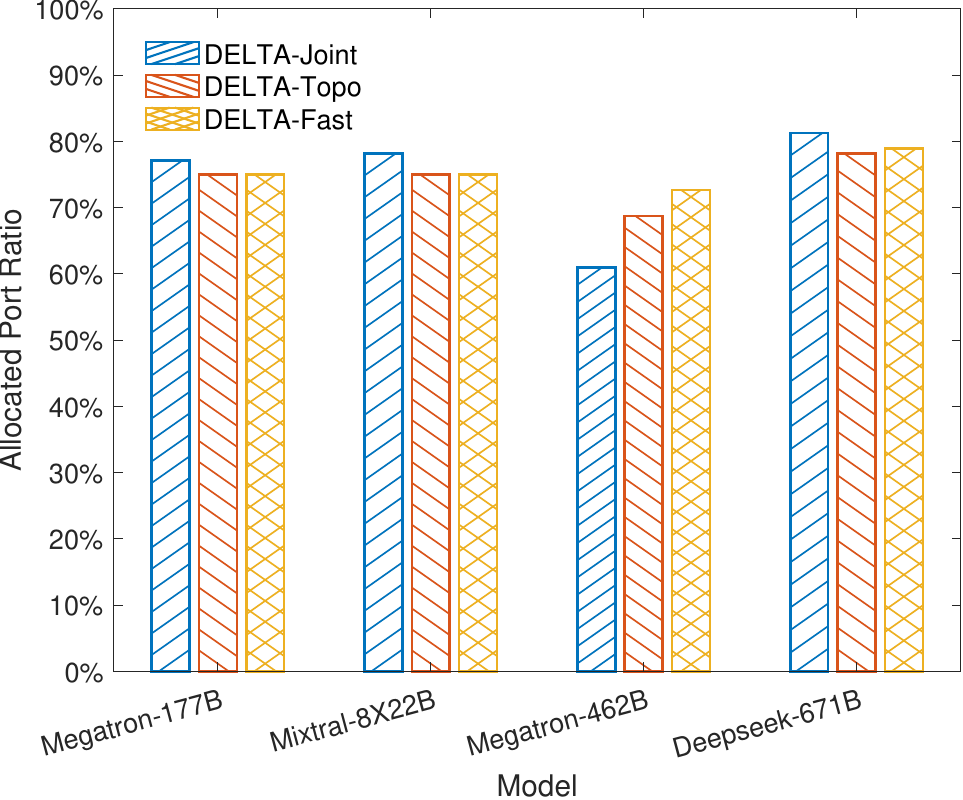}
    \caption{Allocated port ratio compressed by DELTA variants (400 Gb/s, 4096 sequence length).}
    \label{Allocated_Port_Ratio}
\end{figure}

To enhance the performance of bandwidth-bottlenecked workloads, we minimize the total number of allocated optical ports by exploiting temporal slack in communication tasks along non-critical paths, thereby freeing port resources for reallocation. As illustrated in Fig.~\ref{Allocated_Port_Ratio}, this optimization reduces the ratio of allocated ports to available ports to below 80\% (except for DELTA-Joint on DeepSeek-671B at 81.3\%) without prolonging the LLM training time. The port reduction is most pronounced for the Megatron-462B model, which features a higher computation-to-communication ratio in its training; in this scenario, DELTA-Topo and DELTA-Joint further compress the ratios to 68.8\% and 60.9\%, respectively.

\begin{figure}[!h]
    \centering
    \includegraphics[width=0.65\columnwidth]{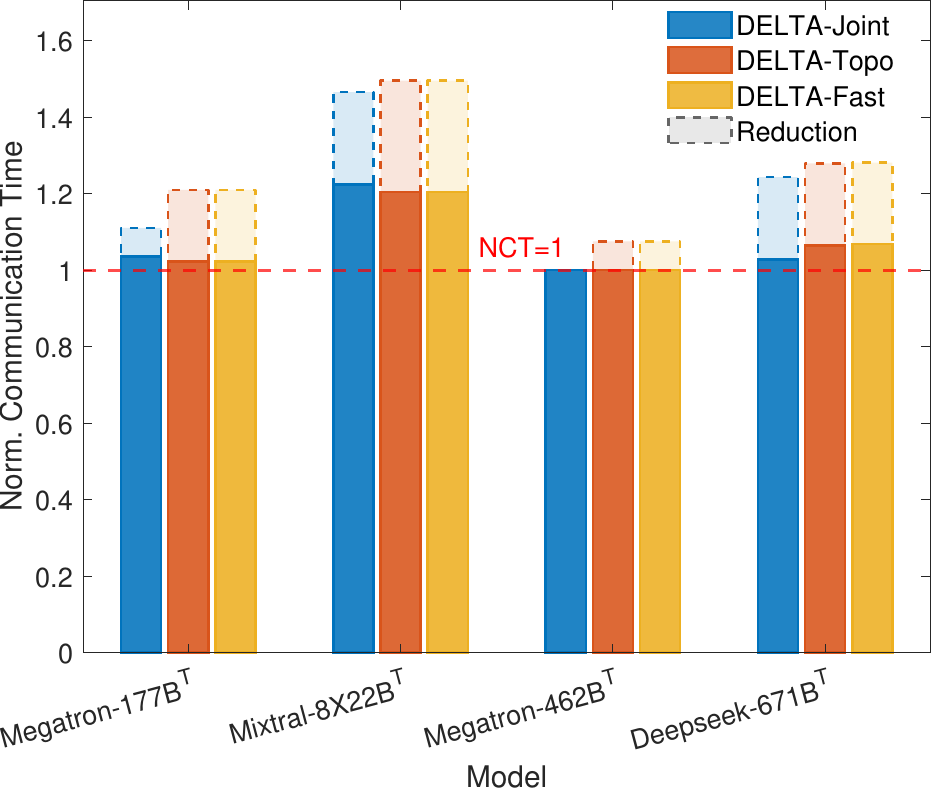}
    \caption{Reduced NCTs of bandwidth-bottlenecked workloads by reallocating surplus ports (400 Gb/s, 4096 sequence length).}
    \label{Improved_NCT}
\end{figure}

Fig.~\ref{Improved_NCT} illustrates the reduced NCTs achieved by reallocating these recovered surplus ports to bandwidth-bottlenecked workloads. Specifically, $\text{Model}^T$ is evaluated by deploying it with a reversed stage-to-pod mapping relative to the original $\text{Model}$ to absorb its released ports (serving as a controlled evaluation scheme to isolate performance gains from broader multi-tenant co-scheduling complexities that lie beyond the scope of this work). Benchmarked against an ideal non-blocking electrical switch operating at 400 Gb/s with a sequence length of 4096, this reallocation strategy significantly narrows the performance gaps for bandwidth-bottlenecked workloads. Specifically, the NCT overheads for Megatron-177B, Mixtral-8X22B, and DeepSeek-671B are substantially reduced from 11.0\% to 2.2\%, 46.5\% to 20.4\%, and 24.3\% to 2.8\%, respectively. This degree of optimization is unattainable for conventional algorithms that construct the logical topology based solely on traffic matrices. Moreover, it highlights a promising strategy for multi-tenant AIDCs: when coupled with an appropriate job placement strategy, compressing the port allocation of bandwidth-insensitive workloads (e.g., Megatron-462B)--without inflating their iteration times--can be leveraged to provision bandwidth-bottlenecked workloads, thereby enabling otherwise severely constrained workloads to approach the performance of ideal non-blocking electrical networks. Notably, for Megatron-177B and Mixtral-8X22B, DELTA-Joint yields a slightly higher NCT than DELTA-Topo because it releases fewer ports to preserve its original optimal performance; we leave the investigation of these complex trade-offs to future work.

\subsection{Execution Time Analysis}

As detailed in Section \ref{acc_MILP}, the algorithm's computational complexity is dominated by the total number of inter-pod communication tasks, denoted as $|\mathcal{M}|$. The scale of $|\mathcal{M}|$ depends on two parameters: the PP size, which dictates the number of stages and proportionally scales both PP and DP task counts; and the number of micro-batches (hereafter, \# of MBS), which scales the number of PP tasks linearly\footnote{Specifically, under the 1F1B scheduling scheme, after aggregating the communication tasks that concurrently start and finish at each stage, the number of PP tasks per model replica is $2 \times (\text{PP size} - 1) \times (\text{\# of MBS})$, while the number of DP tasks equals the PP size.}. Since the configured PP size in Table \ref{tab_llm_workloads} already reaches the recommended upper bound in the literature \cite{liu_moe_2025, nvidiacorporation_performance_2025, nvidiacorporation_megatronlm_2026}, and further scaling PP size is limited by pipeline bubble overheads, this evaluation isolates the impact of scaling the \# of MBS on execution time.

\begin{figure}[!h]
    \centering
    \begin{subfigure}[c]{0.24\textwidth}
        \includegraphics[width=\linewidth]{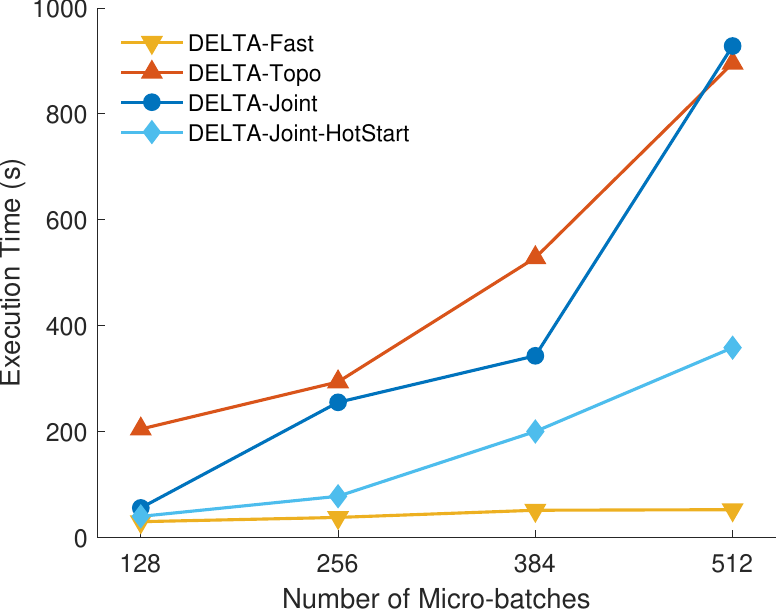}
        \caption{Megatron-462B.}
        \label{Megatron_462B_Time}
    \end{subfigure}
    \begin{subfigure}[c]{0.24\textwidth}
        \includegraphics[width=\linewidth]{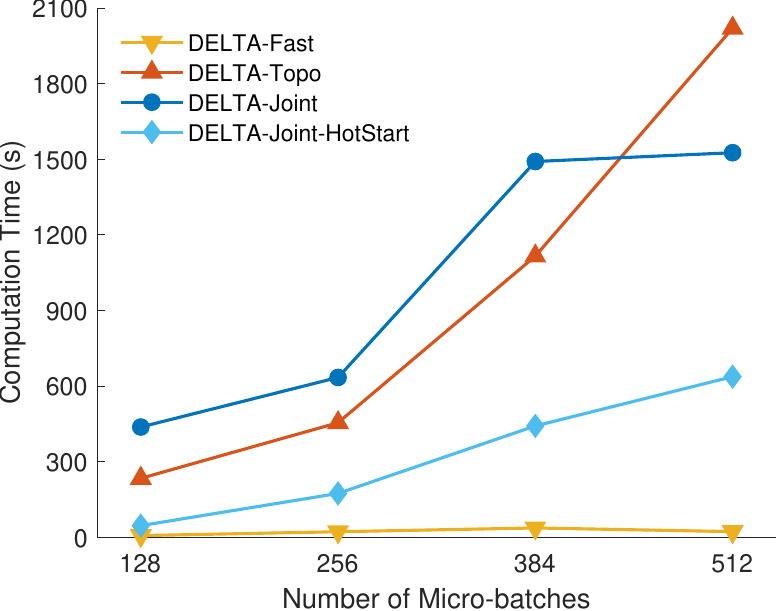}
        \caption{DeepSeek-671B.}
        \label{Deepseek_671B_Time}
    \end{subfigure}
    \caption{Execution time of the algorithms in DELTA.}
    \label{Computation_Time}
\end{figure}

Fig.~\ref{Computation_Time} presents the execution times (including preprocessing time) of DELTA-Fast, DELTA-Topo, DELTA-Joint, and DELTA-Joint-HotStart while varying the \# of MBS when optimizing topologies for Megatron-462B and DeepSeek-671B. Notably, 512 is the maximum configuration reported in \cite{nvidiacorporation_performance_2025}, whereas typical deployments adopt 128 or 256. Benefiting from the acceleration strategies in Section \ref{sim_base_GA}, DELTA-Fast's execution time exhibits only marginal growth as the \# of MBS increases. Across all evaluations, it converges to a stable state (where the optimal solution remains unchanged for 200 consecutive iterations) within tens of seconds. In contrast, the time required for DELTA-Topo and DELTA-Joint to converge to the final solution (defined as the point where, after stabilization, the objective gap between consecutive feasible solutions remains below 0.01\%) increases significantly with the \# of MBS. The maximum execution times for the two algorithms reach 896 s and 928 s for Megatron-462B, and 2020 s and 1526 s for DeepSeek-671B, respectively. However, with the strategies in Section \ref{acc_MILP}, the solving process can complete within 600 s for a \# of MBS up to 256. The hot-start mechanism in DELTA-Joint-HotStart substantially reduces overall execution time (DELTA-Topo-HotStart is omitted since DELTA-Fast already yields a near-optimal solution). Specifically, hot-starting reduces this time by an average of 50.4\% (up to 69.6\%) for Megatron-462B and 72.7\% (up to 89.5\%) for DeepSeek-671B, bringing the maximum execution times of DELTA-Joint-HotStart down to 358 s and 637 s for the two models, respectively. Given that these evaluations were performed on a consumer-grade laptop, the results demonstrate the computational efficiency of the proposed algorithms. Furthermore, considering the highly parallelizable nature of our approach, deploying it on multi-core production servers would unlock even greater speed.

\section{Conclusion}

We propose DELTA, a computation-communication DAG-aware, efficient logical topology optimization framework for OCS-AIDCs. By encoding dynamic LLM training traffic features into the optimization model via the DAG and combining this encoding with variable-length time interval MILP modeling, search space pruning, and heuristic acceleration, DELTA generates a high-quality logical topology within minutes. Furthermore, by exploiting temporal slack in non-critical communication tasks, DELTA can reallocate optical ports without compromising iteration time, effectively boosting bandwidth-bottlenecked workloads.

In multi-job scenarios, the disparate iteration cycles of various workloads cause their traffic overlaps to shift continuously, making accurate traffic prediction difficult. Moreover, since a single makespan objective cannot balance cluster throughput and fairness, a pragmatic solution to this complexity is to independently optimize the topology for each job using DELTA and aggregate these decisions via job-specific weights. Consequently, we are currently finalizing a comprehensive framework for the dynamic orchestration of these weights to achieve data-center-level operational optimality.


\section*{Acknowledgment}
This work was supported in part by the National Key Research and Development Project of China under Grant 2024YFB2908301 and in part by the National Natural Science Foundation of China (NSFC) under Grant 62331017. (Corresponding author: Weiqiang Sun.)

{
\printbibliography
}

\appendices

\section{MILP Formulation with Fixed-Time-Step for Logical Topology Optimization}
\label{app_FTS_MILP}

As a comparative baseline for the variable-length time interval MILP formulation, this appendix details the fixed-time-step counterpart, which formulates the topology optimization problem by discretizing the time horizon into uniform slices.

\vspace{1ex}
\noindent\textbf{Temporal Discretization:}

The considered time horizon is partitioned into a set of fixed-duration time slices, indexed by $\mathcal{T} = \{1, 2, \dots, T\}$, where the integer $T$ denotes the total number of slices and each slice has a constant length $\Delta t$.

\vspace{1ex}
\noindent\textbf{Decision Variables:}
\begin{itemize} [leftmargin=*, labelsep=5pt]
    \item $r_{m,t}$: A continuous variable denoting the instantaneous transmission rate of task $m$ during time slice $t$.
    \item $y_{m,t}$: A binary variable indicating whether task $m$ is active during time slice $t$.
    \item $S_{m,t}, C_{m,t}$: Binary auxiliary variables indicating the initiation and completion of task $m$ at time slice $t$.
    \item $u_{i,j,t}$: A continuous auxiliary variable denoting the normalized fair-share reference rate across the aggregated optical circuits between pod $i$ and pod $j$ during time slice $t$. It facilitates the fair distribution (optional) of aggregated bandwidth across all active tasks sharing this inter-pod connection.
    \item $x_{i,j}$: An integer variable denoting the number of optical circuits/ports allocated between pod $i$ and pod $j$.
\end{itemize}

Unless otherwise specified, all indices $i, j$ range over $\mathcal{P}$, $g$ ranges over $\mathcal{G}$, $t$ ranges over $\mathcal{T}$, and $m$ belongs to $\mathcal{M}$.

\vspace{1ex}
\noindent\textbf{Objectives:}

The primary optimization objective is to minimize the iteration time of the training task:
\begin{equation}
    \min \quad C.
    \label{eq:app_primary_obj}
\end{equation}

A secondary objective minimizes total optical port consumption without increasing the optimal makespan $C^*$:
\begin{equation}
    \min \quad \sum_{i \in \mathcal{P}} \sum_{j \in \mathcal{P}, j \neq i} x_{i,j}, \quad \text{s.t.} \quad C \le C^*.
\label{eq:app_secondary_obj}
\end{equation}

\vspace{1ex}
\noindent\textbf{Constraints:}

\subsubsection*{1) Topology-Related Constraints}
\begin{equation}
\begin{cases}
\sum_{j \in \mathcal{P}, j \neq i} x_{i,j} \le U_i, & \forall i. \\
\sum_{i \in \mathcal{P}, i \neq j} x_{i,j} \le U_j, & \forall j. \\
x_{i,j} = x_{j,i}, & \forall i, j.
\end{cases}
\label{eq:app_topology}
\end{equation}

Eq.~\eqref{eq:app_topology} defines the physical port limits and bidirectional symmetry, following the same logic as the main text.

\subsubsection*{2) Optical Circuit and NIC Capacity-Related Constraints}
\begin{equation}
    \sum_{m \in \mathcal{M}_{(i,j)}} r_{m,t} \le x_{i,j} \cdot B, \quad \forall t, \forall i, j.
    \label{eq:app_link_capacity}
\end{equation}

\begin{equation}
    \begin{cases} 
    \sum_{m \in \Phi_{\text{src}}(g)} \frac{r_{m,t}}{F_m} \le B, & \forall t, g. \\ 
    \sum_{m \in \Phi_{\text{dst}}(g)} \frac{r_{m,t}}{F_m} \le B, & \forall t, g. 
    \end{cases}
    \label{eq:app_nic_capacity}
\end{equation}
Eqs.~\eqref{eq:app_link_capacity} and \eqref{eq:app_nic_capacity} bound the aggregated rates by the allocated logical link bandwidth and individual NIC injection/reception capacities, respectively.

\subsubsection*{3) Task Lifecycle and Data Transmission Constraints}

\begin{equation}
    \sum_{t} S_{m,t} = 1, \quad \sum_{t} C_{m,t} = 1, \quad \forall m.
    \label{eq:app_task_lifecycle}
\end{equation}

\begin{equation}
    y_{m,t} - y_{m,t-1} = S_{m,t} - C_{m,t}, \quad \forall m, t.
    \label{eq:app_state_continuity}
\end{equation}

\begin{equation}
    \sum_{t=1}^{T} (r_{m,t} \cdot \Delta t) \ge V_m, \quad \forall m.
    \label{eq:app_data_conservation}
\end{equation}

\begin{equation}
    r_{m,t} \le y_{m,t} \cdot (F_m \cdot B), \quad \forall m, t.
    \label{eq:app_rate_mapping}
\end{equation}

Eqs.~\eqref{eq:app_task_lifecycle} and \eqref{eq:app_state_continuity} enforce unique initiation and completion points, maintaining state continuity throughout the lifecycle of each task. Eq.~\eqref{eq:app_data_conservation} ensures that the cumulative data volume transmitted satisfies the predefined demand $V_m$. Furthermore, Eq.~\eqref{eq:app_rate_mapping} explicitly couples the continuous transmission rate with the binary state, ensuring that transmission only occurs during active intervals ($y_{m,t} = 1$) while respecting the flow-level physical bandwidth limits.

\subsubsection*{4) Inter-pod Communication DAG Constraints}
\begin{equation}
\begin{aligned}
& \sum_{t} (t \cdot S_{m, t}) \ge \sum_{t} (t \cdot C_{m_{\text{pre}}, t}) + \left\lceil \frac{\delta_{m_{\text{pre}} \to m}}{\Delta t} \right\rceil, \\
& \forall (m_{\text{pre}}, m, \delta) \in \mathcal{D}.
\end{aligned}
\label{eq:app_dag_constraint}
\end{equation}
Eq.~\eqref{eq:app_dag_constraint} ensures that task execution strictly adheres to the precedence requirements defined in the reduced DAG $\mathcal{D}$.

\subsubsection*{5) Flow Fairness Constraints (Optional for Comparison)}
\begin{equation}
\begin{aligned}
& -M(1-y_{m,t}) \le \frac{r_{m,t}}{F_m} - u_{i,j,t} \le M(1-y_{m,t}), \\
& \forall t, (i,j), m \in \mathcal{M}_{(i,j)}.
\end{aligned}
\label{eq:app_fairness}
\end{equation}
Eq.~\eqref{eq:app_fairness} uses the Big-M method to simulate conventional fair-sharing mechanisms by equalizing the normalized rates of concurrent flows on the same link.

\subsubsection*{6) Objective Function Constraint}
\begin{equation}
    C \ge (t \cdot \Delta t) C_{m,t}, \quad \forall m, t.
    \label{eq:app_makespan_def}
\end{equation}
Eq.~\eqref{eq:app_makespan_def} defines the global completion time as the envelope of all individual task completion times.

\section{Limitations of Topological-Sorting-Based Search Space Pruning} \label{app_limitations}

Relying exclusively on topological sorting for search space pruning is conservative. First, when two tasks (e.g., the B3S2 PPFwd and B1S3 PPBwd nodes in Fig.~\ref{Dependency_Graph}) share the same topological order on $\mathcal{D}$ but exhibit different time intervals $\delta$ relative to their predecessors, the task subject to a longer delay should logically be activated in a later time interval. However, because its earliest active index remains identical to that of the earlier task, topological sorting yields lower-bound (left-side) index redundancy (as shown by the left boundary of the feasible region for the fourth task in Fig.~\ref{retained_interval}).

Second, consider a task serving as a common predecessor across multiple dependency branches (e.g., the B4S3 PPBwd node in Fig.~\ref{Dependency_Graph}, which is a prerequisite for three different DP communication tasks). If its multiple successors have varying startup interval requirements, this predecessor should logically complete earlier to accommodate their distinct active intervals. Yet, when determining its latest active index, topological sorting adopts the pessimistic assumption that all successors might start and stop simultaneously (i.e., collectively occupying only a single time interval). This oversimplification generates upper-bound (right-side) index redundancy (as demonstrated by the right boundary of the feasible region for the fourth-to-last task in Fig.~\ref{retained_interval}).

\section{Supplementary Algorithms for Search Space Pruning}
\label{app_search_space_pruning}

\begin{figure}[!h]
    \centering
    \includegraphics[width=0.6\columnwidth]{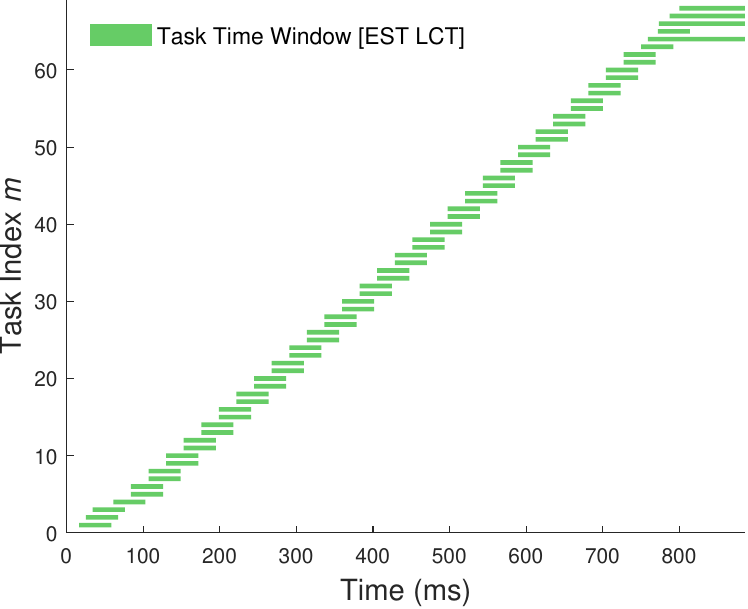}
    \caption{DAG-aware task time window estimation for search space pruning (LLM training setup as in Fig.~\ref{task_interval_window}).}
    \label{task_time_window}
\end{figure}

This appendix details Alg.~\ref{alg_cal_task_time_windows} (\texttt{CalTaskTimeWindows}), which is used in Alg.~\ref{alg_capacity_bound}. Alg.~\ref{alg_cal_task_time_windows} computes the earliest start time (EST) and latest completion time (LCT) for each task $m \in \mathcal{M}$. The algorithm operates in three steps. First, it initializes the minimum physical duration $\tau_m$ based on the task data volume $V_m$ and per-flow bandwidth $B$ (Line 2). Second, a forward propagation traversal over the reduced computation-communication DAG $\mathcal{D}$ calculates the EST for each task by respecting all rigid dependency intervals $\delta$ (Lines 6--13). Finally, backward propagation starting from the coarsely estimated upper bound on the iteration time $\hat{T}_{up}$ yields the LCT (Lines 14--21). The effect of Alg.~\ref{alg_cal_task_time_windows} is illustrated in Fig.~\ref{task_time_window}.

\begin{algorithm}[htbp]
\caption{CalTaskTimeWindows}
\label{alg_cal_task_time_windows}
\begin{algorithmic}[1]
\Require $\mathcal{M}$, $\mathcal{D}$, bandwidth $B$, time upper bound $\hat{T}_{up}$.
\Ensure $\text{EST}, \text{LCT}$.

\vspace{0.5em}
\Statex \hspace{-\algorithmicindent} \textbf{$\triangleright$ Step 1: Initialization \& Minimum Duration Estimation}
\For{\textbf{each} $m \in \mathcal{M}$}
    \State $\tau_m \gets V_m / (F_m \cdot B)$ \Comment{Minimum physical duration}
    \State $\text{EST}_m \gets 0, \ \text{LCT}_m \gets \hat{T}_{up}$
    \State Compute in-degree $deg_{\text{in}}(m)$ and out-degree $deg_{\text{out}}(m)$ based on $\mathcal{D}$.
\EndFor

\vspace{0.5em}
\Statex \hspace{-\algorithmicindent} \textbf{$\triangleright$ Step 2: Forward Propagation (Earliest Start Time)}
\State $\mathcal{Q} \gets \{m \in \mathcal{M} \mid deg_{\text{in}}(m) = 0\}$
\While{$\mathcal{Q} \neq \emptyset$}
    \State $u \gets \mathcal{Q}.\text{pop}()$
    \For{\textbf{each} dependency $(u, v, \delta) \in \mathcal{D}$}
        \State $\text{EST}_v \gets \max(\text{EST}_v, \text{EST}_u + \tau_u + \delta)$
        \State $deg_{\text{in}}(v) \gets deg_{\text{in}}(v) - 1$; \textbf{if} $deg_{\text{in}}(v) == 0$ \textbf{then} $\mathcal{Q}.\text{push}(v)$
    \EndFor
\EndWhile

\vspace{0.5em}
\Statex \hspace{-\algorithmicindent} \textbf{$\triangleright$ Step 3: Backward Propagation (Latest Completion Time)}
\State $\mathcal{Q} \gets \{m \in \mathcal{M} \mid deg_{\text{out}}(m) = 0\}$
\While{$\mathcal{Q} \neq \emptyset$}
    \State $v \gets \mathcal{Q}.\text{pop}()$
    \For{\textbf{each} dependency $(u, v, \delta) \in \mathcal{D}$}
        \State $\text{LCT}_u \gets \min(\text{LCT}_u, \text{LCT}_v - \tau_v - \delta)$
        \State $deg_{\text{out}}(u) \gets deg_{\text{out}}(u) - 1$; \textbf{if} $deg_{\text{out}}(u) == 0$ \textbf{then} $\mathcal{Q}.\text{push}(u)$
    \EndFor
\EndWhile
\State \Return $\text{EST}, \text{LCT} = \{\text{EST}_m, \text{LCT}_m \mid m \in \mathcal{M}\}$

\end{algorithmic}
\end{algorithm}

Note that Fig.~\ref{task_time_window} and Fig.~\ref{retained_interval} exhibit similarly compact profiles, suggesting that an analogous approach could be employed to prune the search space for the fixed-time-step MILP formulation. However, in our empirical evaluations, even with a relatively coarse time resolution of 0.1 ms (compared to the millisecond-level duration of a single PP communication) and similar search space pruning techniques, solving the fixed-time-step MILP corresponding to a moderately large-scale problem still required tens of hours.

\section{Subroutines for Domain-Adapted Genetic Algorithm}
\label{app_ga_subroutines}

This appendix details the initialization and repair subroutines used by the domain-adapted genetic algorithm in Alg.~\ref{alg_domain_adapted_genetic}. Alg.~\ref{alg_feasible_random_init} (\texttt{FeasibleRandomInit}) constructs a valid initial logical topology $X$ that adheres to both the physical port capacities $U$ and the upper bounds $\bar{X}$. First, it initializes the current port usage and calculates the unprocessed degrees for all pods to anticipate future connectivity requirements (Lines 1--2). Second, it iteratively processes each active pod pair $e \in E$ to determine the maximum allowable capacity by reserving ports for remaining connections and bounding it with $\bar{X}_e$ (Lines 3--10). Finally, it uniformly samples a valid link capacity $x_e$ within the calculated limits and updates the physical port usage for the corresponding endpoints (Lines 11--15).

\begin{algorithm}[htbp]
\caption{FeasibleRandomInit}
\label{alg_feasible_random_init}
\begin{algorithmic}[1]
\Require Active pod pairs (edges) $E$, port capacities $U=[U_1,...,U_{|\mathcal{P}|}]$, upper bounds $\bar{X}$
\Ensure A feasible logical topology $X$

\State Initialize port usage $U^{\text{used}}_p \gets 0$ for all $p \in \mathcal{P}$

\vspace{0.5em}
\Statex \hspace{-\algorithmicindent} \textbf{$\triangleright$ Initialize unprocessed degrees for future lookahead}
\State Let $D_p \gets |\{e \in E \mid e \text{ connects to } p\}|$ for all $p \in \mathcal{P}$

\For{\textbf{each} edge $e = (u, v) \in E$}
    \State \textbf{$\triangleright$ Update future degrees (remaining ports after current $e$)}
    \State $D_u \gets D_u - 1$; \quad $D_v \gets D_v - 1$

    \State \textbf{$\triangleright$ Calculate remaining valid ports}
    \State $R_u \gets U_u - U^{\text{used}}_u$; \quad $R_v \gets U_v - U^{\text{used}}_v$

    \State \textbf{$\triangleright$ Reserve ports for future connectivity}
    \State $R'_u \gets R_u - D_u$; \quad $R'_v \gets R_v - D_v$

    \State \textbf{$\triangleright$ Determine maximum allowable capacity and bound it}
    \State $limit \gets \min(R'_u, R'_v, \bar{X}_{e})$
    \State $limit \gets \max(1, limit)$ \Comment{Ensure basic connectivity}

    \State \textbf{$\triangleright$ Sample capacity and update usage}
    \State $x_{e} \sim \mathcal{U}\{1, limit\}$
    \State $U^{\text{used}}_u \gets U^{\text{used}}_u + x_{e}$; \quad $U^{\text{used}}_v \gets U^{\text{used}}_v + x_{e}$
\EndFor

\State \Return $X = \{x_{e} \mid e \in E\}$
\end{algorithmic}
\end{algorithm}
\vspace{-0em}
To address invalid offspring generated during crossover and mutation, Alg.~\ref{alg_repair_topo} (\texttt{RepairTopo}) restores compliance with all physical and logical constraints. First, a base trimming process enforces the capacity upper bounds $\bar{X}_e$ and ensures connectivity for each active pod pair (Lines 1--3). Second, a port overflow reduction mechanism identifies overloaded pods and randomly decrements the capacity of their reducible edges until the physical port limits are met or no reducible edges remain (Lines 4--15). Finally, a verification step checks whether all port constraints $U_p$ are satisfied, and returns the repaired topology and a boolean success flag (Lines 16--20).

\begin{algorithm}[!h]
\caption{RepairTopo}
\label{alg_repair_topo}
\begin{algorithmic}[1]
\Require Mutant/crossover topology $X'$, active pod pairs $E$, port capacities $U=[U_1,...,U_{|\mathcal{P}|}]$, upper bounds $\bar{X}$
\Ensure Repaired topology $X'$, boolean flag $success$ indicating repair status

\vspace{0.5em}
\Statex \hspace{-\algorithmicindent} \textbf{$\triangleright$ Step 1: Base Trimming}
\For{\textbf{each} $e \in E$}
    \State $x'_{e} \gets \max\big(1, \min(x'_{e}, \bar{X}_{e})\big)$ \Comment{Enforce bounds \& connectivity}
\EndFor

\vspace{0.5em}
\Statex \hspace{-\algorithmicindent} \textbf{$\triangleright$ Step 2: Port Overflow Reduction}
\State Compute current port usage $U^{\text{used}}_p$ for all $p \in \mathcal{P}$
\While{$\exists p \in \mathcal{P} \text{ such that } U^{\text{used}}_p > U_p$}
    \State Let $\mathcal{P}_{\text{over}}$ be the set of overloaded pods
    \State Randomly select a pod $p \in \mathcal{P}_{\text{over}}$
    \State $E_p \gets \{e \in E \mid e \text{ connects to } p \text{ and } x'_e > 1\}$

    \If{$E_p = \emptyset$}
        \State \textbf{break} \Comment{Repair failed: overloaded but no reducible edges}
    \EndIf

    \State Randomly select a reducible edge $e \in E_p$
    \State $x'_e \gets x'_e - 1$
    \State Update $U^{\text{used}}$ locally for the endpoints of $e$
\EndWhile

\vspace{0.5em}
\Statex \hspace{-\algorithmicindent} \textbf{$\triangleright$ Step 3: Final Verification}
\If{all $U^{\text{used}}_p \le U_p$ \textbf{for} $p \in \mathcal{P}$}
    \State \Return $X', \text{True}$
\Else
    \State \Return $X', \text{False}$
\EndIf

\end{algorithmic}
\end{algorithm}
\vfill


\end{document}